%% file: main.tex
\documentclass[twocolumn]{aastex631}

\newcommand\latex{La\TeX}

\newcommand{\nustar}{\rm{NuSTAR}}
\newcommand{\rxte}{\rm{RXTE}}
\newcommand{\suzaku}{\rm{Suzaku}}
\newcommand{\xmm}{\rm{XMM-Newton}}
\newcommand{\chandra}{\rm{Chandra}}
\newcommand{\ginga}{\rm{Ginga}}
\newcommand{\bepposax}{\rm{BeppoSAX}}
\newcommand{\uhuru}{\rm{Uhuru}}

\newcommand{\nhim}{N_{\rm H}^{\rm IM}}
\newcommand{\nh}{N_{\rm H}}

\usepackage{comment}
\usepackage{multirow}

\submitjournal{ApJ}

\shorttitle{Cen X-3 NuSTAR}
\shortauthors{Tamba et al.}
\graphicspath{{./}{figures/}}

\newcommand{\red}[1]{{#1}}

\newcommand{\redunderline}[1]{{#1}}

\begin{document}

\title{Orbital- and spin-phase variability in the X-ray emission from the accreting pulsar Cen X-3}

\correspondingauthor{Tsubasa Tamba}
\email{tsubasa.tamba@phys.s.u-tokyo.ac.jp}

\author[0000-0001-7631-4362]{Tsubasa Tamba}
\affiliation{Department of Physics, Faculty of Science, The University of Tokyo, 7-3-1 Hongo, Bunkyo-ku, Tokyo 113-0033, Japan}

\author[0000-0003-2670-6936]{Hirokazu Odaka}
\affiliation{Department of Physics, Faculty of Science, The University of Tokyo, 7-3-1 Hongo, Bunkyo-ku, Tokyo 113-0033, Japan}
\affiliation{Kavli IPMU, The University of Tokyo, Kashiwa 113-0033, Japan}
\affiliation{Research Center for Early Universe, Faculty of Science, The University of Tokyo, 7-3-1 Hongo, Bunkyo-ku, Tokyo 113-0033, Japan}

\author[0000-0002-0114-5581]{Atsushi Tanimoto}
\affiliation{Graduate School of Science and Engineering, Kagoshima University, 1-21-24, Korimoto, Kagoshima, Kagoshima 890-0065, Japan}

\author[0000-0003-4237-1101]{Hiromasa Suzuki}
\affiliation{Department of Physics, Konan University, 8-9-1 Okamoto, Higashinada-ku, Kobe, Hyogo, Japan}

\author[0000-0003-0590-6330]{Satoshi Takashima}
\affiliation{Department of Physics, Faculty of Science, The University of Tokyo, 7-3-1 Hongo, Bunkyo-ku, Tokyo 113-0033, Japan}

\author[0000-0003-0890-4920]{Aya Bamba}
\affiliation{Department of Physics, Faculty of Science, The University of Tokyo, 7-3-1 Hongo, Bunkyo-ku, Tokyo 113-0033, Japan}
\affiliation{Research Center for Early Universe, Faculty of Science, The University of Tokyo, 7-3-1 Hongo, Bunkyo-ku, Tokyo 113-0033, Japan}

\begin{abstract}

\input{abstract.tex}

\end{abstract}

\input{section1.tex}

\input{section2.tex}

\input{section3.tex}

\input{section4.tex}

\input{section5.tex}

\input{section6.tex}

\input{section7.tex}

\begin{acknowledgments}
We thank the anonymous referee, who improved this work with his or her valuable comments.
This work is supported by the Japan Society for the Promotion of Science (JSPS) Research Fellowship for Young Scientist No. 20J20050 (TT), JSPS/MEXT KAKENHI grant numbers 18H05861, 19H01906, 19H05185, 22H00128, 22K18277 (HO), 20J00119 (AT), 21J00031 (HS), and 19K03908 (AB).
\end{acknowledgments}

%

\vspace{5mm}
\facilities{NuSTAR \citep{Harrison2013}}


\software{HEASoft, Xronos, XSPEC \citep{Arnaud1996}, matplotlib \citep{Hunter2007}
          }

\bibliography{reference}{}
\bibliographystyle{aasjournal}


\end{document}

%% file: abstract.tex
We present a time-resolved analysis on 39 ks NuSTAR observation data of the X-ray pulsar Centaurus~X-3 (Cen~X-3), which covered an orbital-phase interval of $\Phi=0.199$--$0.414$.
The orbital- and spin-phase variability was investigated through time-resolved spectra, light curves, and pulse profiles.
The orbital-phase variability was due to the mixture of two comparable effects, the intrinsic flux variability of $\sim10\%$ and the obscuration by the clumpy stellar wind.
The typical size and number density of the clumps are $\sim9\times10^{10}\;{\rm cm}$ and $\sim3\times10^{12}\;{\rm cm^{-3}}$, respectively.
In the spin-phase-resolved analysis, we detected variations in the spectral features of the continuum, the Fe line, and the cyclotron resonance scattering feature (CRSF).
The photon index ranged from 0.72 to 1.06, corresponding to the difference in the Comptonization optical depth by a factor of $\sim1.6$.
The equivalent width and intensity of the Fe line had negative correlations with the continuum flux.
The central energy and the strength of the CRSF increased at the pulse maximum.
The former ranged from 26.0 to 28.7 keV, and the latter varied by a factor of $\sim1.9$.
The pulse profile was double-peaked in the low-energy band and gradually shifted to single-peaked with energy, indicating the existence of two distinct emission patterns corresponding to the pencil and fan beams.
Finally, we found that the pulse profiles were highly stable along the orbital phase within a variation degree of $\sim20\%$, which gives evidence of a highly stable accretion stream of the binary system.

%% file: section1.tex
\section{Introduction} \label{sec:intro}

X-ray pulsars, binary systems of a neutron star and a massive star, constitute one of the brightest X-ray source categories in our galaxy that emit X-rays via accretion of gas provided by their donor stars.
It is usually accompanied by strong magnetic fields of $B\sim10^{12}\;{\rm G}$ \citep[for a recent review, see][]{Mushtukov2022}.
Such objects provide us with ideal and unique laboratories to study structures of accretion flows, physical processes under strong magnetic field, and properties of the stellar winds from donor stars \citep[for a review, see][]{Martinez2017}.

\red{Many physical models have been constructed to explain the emission mechanism of X-ray pulsars.}
The most prevailing theory is the accretion column model \citep{Basko1976, Becker1998}.
In this model, the accretion flow is channeled along the magnetic field lines onto the magnetic poles, leading to the formation of optically thick accretion columns just above the poles.
\red{When the luminosity exceeds the critical value of $\sim10^{37}\;{\rm erg\;s^{-1}}$ \citep{Mushtukov2015},} 
the radiation-dominated shock inside the column decelerates the accretion flow, and the kinetic energy is converted to X-rays via Comptonization of seed photons emitted from the surface of the neutron star \citep{Becker2005, Becker2007, Odaka2014}.
The power-law-like spectrum accompanied by a cut-off feature, typical of X-ray pulsars, is interpreted as a consequence of bulk and thermal Comptonization in the accretion column with substantial modification of scattering cross sections due to the presence of the magnetic field \redunderline{\red{\citep{Daugherty1986, Mushtukov2016, Schwarm2017, Suleimanov2022}}}.
Some previous studies have revealed that the model satisfactorily explains the X-ray spectra of X-ray pulsars \citep[e.g.,][]{Wolff2016, West2017} based on the one-dimensional radiative transfer equation.
\red{Other geometries or emission processes have been proposed for the X-ray pulsars with sub-critical luminosities.
One possibility is the deceleration of the accretion stream by the collisionless shock on the top of the column \citep{Langer1982, Bykov2004}.
The radiation from hot spots near the neutron star surface \citep{Zeldovich1969} is another possibility, which gives a different emission geometry and spectral formation from the accretion column model \citep{Mushtukov2021, Sokolova-Lapa2021}.}

\red{The understanding of the emission spectra based on three-dimensional structures is still underway, such as binary motion, rotation of the neutron star, and geometrical relation among the spin axis, the magnetic pole, and the line of sight.
Moreover, the surrounding environment, such as stellar wind and accretion disk, is an essential factor that affects the observed spectrum, but little has been understood about their geometrical structures.
These three-dimensional structures can be investigated by examining the spectral variabilities of X-ray pulsars.
They are consequences of complicated mixtures of the variability of the accretion rate, the difference in physical processes depending on emission directions, and the obscuration by an inhomogeneous stellar wind.}
A good example is Vela~X-1, the brightest wind-fed X-ray pulsar, showing high variability.
The spin-phase-averaged variability in luminosity is explained by the mass capture rate of inhomogeneous stellar wind \citep{Kreykenbohm2008}.
This change in the accretion rate induces the time-dependent structure of the accretion column, which results in the spectral variability of the Comptonized radiation from the column \citep{Odaka2013, Odaka2014}.
However, such wind-fed sources make it difficult to investigate the three-dimensional picture by spin-phase-resolved spectroscopy due to the highly transient accretion streams.

Centaurus~X-3 (Cen X-3) is one of the best targets to investigate the relation between spectral variability and physical conditions.
This is because it is a representative disk-fed X-ray pulsar that has a highly stable accretion stream compared to wind-fed sources.
In such an object, it is more straightforward to observationally distinguish the effects of stellar wind and intrinsic accretion stream by applying orbital- and spin-phase-resolved spectroscopy.
The existence of the accretion disk is suggested by the high luminosity of $\sim5.0\times10^{37}\;{\rm erg\;s^{-1}}$ \citep{Suchy2008}, the variability of optical light curve \citep{Tjemkes1986}, and quasi-periodic oscillations at $\sim40\;{\rm mHz}$ \citep{Takeshima1991, Raichur2008_QPO}.
\red{Since the luminosity exceeds the critical value, the radiation-dominated shock inside the accretion columns is considered to be responsible for the X-ray emission.}
The well-known binary properties of Cen~X-3 also make it a suitable target for investigating physical conditions.
It is the first binary pulsar discovered in X-rays \citep{Chodil1967, Giacconi1971}, and one of the brightest X-ray pulsars, located at a distance of $6.4^{+1.0}_{-1.4}\;{\rm kpc}$ from the earth \citep{Arnason2021}.
In this system, a neutron star with a mass of $1.34^{+0.16}_{-0.14}\;M_{\odot}$ \citep{venderMeer2007} is orbiting a giant O6--8 III counterpart V779 Cen with a mass and radius of $20.5\pm0.7\;M_{\odot}$ and $12\;R_{\odot}$ \citep{Schreier1972, Krzeminski1974, Hutchings1979}, respectively.
The orbital period of the system is $\sim 2.08\;{\rm d}$ \citep{Bildsten1997}, and about $\sim 20\%$ of the orbit is eclipsed by the companion star due to the high inclination angle of $70.2\pm2.7\;{\rm deg}$ \citep{Ash1999}.
The neutron star spin period of $\sim4.8\;{\rm s}$ is first discovered by \uhuru\ \citep{Giacconi1971}, followed by the discovery of a gradual spin-up feature as a result of mass accretion \citep[e.g.,][]{Tsunemi1996}.

The spin-phase-averaged spectrum of Cen~X-3 is characterized by a power law with a photon index of $\sim 1.0$ with a high-energy cut-off around $\sim 15\;{\rm keV}$ \citep[e.g.,][]{White1983}, accompanied by several additional components.
One of those components is the cyclotron resonance scattering feature (CRSF) around $30\;{\rm keV}$, which was first reported by \ginga\ \citep{Nagase1992} and later confirmed by \bepposax\ \citep{Santangelo1998}, indicating that the magnetic field at the surface of the neutron star is $B\sim3\times10^{12}\;{\rm G}$.
The spin-phase variability of the CRSF is reported by \cite{Burderi2000}.
Another important spectral component is the Fe K-emission line first observed by \ginga\ \citep{Day1993}.
It is composed of three emission lines at $\sim6.4\;{\rm keV}$, $\sim6.7\;{\rm keV}$, and $\sim7.0\;{\rm keV}$, which were resolved by ASCA \citep{Ebisawa1996}.
The orbital- and spin-phase spectral variabilities of the continuum were also studied with several sets of observation data (\chandra: \citealp{Wojdowski2003, Iaria2005}; \rxte: \citealp{Raichur2008}; \suzaku: \citealp{Naik2011}; \xmm: \citealp{Sanjurjo2021}).
\red{Most recently, the spin-phase variability of the polarization was investigated with IXPE to determine the emission geometry as displaced from antipodal configuration \citep{Tsygankov2022}, consistent with the results of the pulse deconstruction analysis performed by \cite{Kraus1996}.}
However, the physical conditions that cause spectral variability are still under debate, and no consensus has been reached until now about the stability of the accretion stream or the structure of the stellar wind.

In this paper, we aim to interpret the orbital- and spin-phase variability of Cen~X-3 by evaluating the stability of the accretion stream and geometrical structure of the stellar wind.
We investigated soft and hard X-ray variabilities of the source by utilizing a set of $39\;{\rm ks}$ \nustar\ observation data via orbital- and spin-phase-resolved analysis on spectra, light curves, and pulse profiles.
Since the phase-averaged spectrum was analyzed by \cite{Tomar2021} in a phenomenological way and by \cite{Thalhammer2021} via the accretion column model, we focus on the detailed time-resolved analysis of the observation data.
The remainder of this paper is organized as follows.
In Section \ref{sec:observation_and_data_reduction}, we briefly describe the observation and reduction of the \nustar\ data.
In Section \ref{sec:orbital_phase_resolved}, we give results of the orbital-phase-resolved analysis with energy-resolved light curves and orbital-phase-resolved spectra.
We then investigate the spin-phase variability via energy-resolved pulse profiles and spin-phase-resolved spectroscopy in Section \ref{sec:spin_phase_resolved}.
In Section \ref{sec:independence}, we present the results of the doubly-phase-resolved analysis.
Discussions are presented in Section \ref{sec:discussion}, where we give physical interpretations of the analysis results and compare them with previous studies.
Finally, the concluding remarks are given in Section \ref{sec:conclusions}.

%% file: section2.tex
\section{Observation and data reduction} \label{sec:observation_and_data_reduction}

\nustar\ \citep{Harrison2013} observed Cen~X-3 from 2015 November 30 to December 1 with an elapsed time of $38.7\;{\rm ks}$ (ObsID: 30101055002).
We performed data reduction with the standard \nustar\ analysis software contained in {\tt HEASoft} 6.29.
The source regions were defined as circles with a radius of $180\arcsec$ from the source center and background regions as rectangles in the off-source regions.
The net exposure was $21.4\;{\rm ks}$ and $21.6\;{\rm ks}$ for FPMA and FPMB, respectively.
Cleaned event data were extracted after applying the barycentric correction.
All of the temporal and spectral analyses in this paper were carried out with {\tt Xronos} 6.0 and {\tt XSPEC} 12.12.0, respectively, which are also contained in {\tt HEASoft}.
The uncertainties are given at $1\sigma$ confidence levels unless stated otherwise.

Before temporal or spectral analysis of X-ray binaries, it is necessary to correct the arrival times of observation data for the binary motion.
First of all, we divided the whole observation data into 20 pieces of $2\;{\rm ks}$ data set and applied epoch folding search to them to find the spin period of the pulsar at each time interval.
The spin periods obtained from the 20 intervals were clearly affected by Doppler effect that are induced by the orbital motion. In order to correct the effect, we fitted Doppler-shifted spin periods with a sinusoidal function
\begin{eqnarray}
P_{\rm Doppler}(t)=P_{\rm spin}\left\{1-A\sin\left(2\pi\Phi\right)\right\}.
\end{eqnarray}
$\Phi$ denotes the orbital phase of the binary system, in which $0.0\leq\Phi<1.0$ and $\Phi=0.0$ corresponds to the mid-eclipse.
$P_{\rm Doppler}$ and $P_{\rm spin}$ denote the Doppler-shifted and rest-frame spin periods of the pulsar, respectively. The small eccentricity of $e<0.0016$ \citep{Bildsten1997} was ignored here. The amplitude of the sinusoidal function $A$ can be expressed using orbital parameters by
\begin{eqnarray}
A=\frac{2\pi a_{\rm x}\sin i}{cP_{\rm orb}},
\end{eqnarray}
where $P_{\rm orb}$, $a_{\rm x}\sin i$, and $c$ are the orbital period, the projected semi-major axis of the binary system, and the speed of light, respectively. Table \ref{tab:orbital_parameters} shows orbital parameters calculated from the fitting result. $P_{\rm orb}$ was assumed invariant from the previous observation due to small $\dot{P}_{\rm orb}$.
$P_{\rm spin}$ and $a_{\rm x}\sin i$ derived from the observation data are consistent with previous studies \citep{Bildsten1997, Falanga2015}. The orbital phase at the start time of the observation was $\Phi_{\rm start}=0.199$, which means the $38.7\;{\rm ks}$ \nustar\ observation covers $\Phi=0.199$--$0.414$.

The arrival times of the observed photons can be corrected for the binary motion using the orbital parameters listed in Table \ref{tab:orbital_parameters} as
\begin{eqnarray}
t_{\rm binary}=t_{\rm solar}-\frac{a_{\rm x}\sin i}{c}\cos\left(2\pi\Phi\right),
\end{eqnarray}
where $t_{\rm solar}$ and $t_{\rm binary}$ are the arrival time after the barycentric correction and after the binary correction, respectively.
After the extraction of barycentric corrected event data, we applied the binary correction to the arrival time of each event. All of the temporal analyses in this paper are based on the binary corrected arrival times ($t_{\rm binary}$).

\begin{deluxetable*}{ccccc}
\tablecaption{Orbital parameters of Cen~X-3.\label{tab:orbital_parameters}}
\tablewidth{0pt}
\tablehead{
\colhead{Reference} & \colhead{$P_{\rm orb}$} & \colhead{$P_{\rm spin}$} & \colhead{$a_{\rm x}\sin{i}$} & \colhead{$\Phi_{\rm start}$$^{a}$}\\
& \colhead{(days)} & \colhead{(s)} & \colhead{(lt-s)} & \colhead{}
}
\startdata
This work & $2.087$ (fixed) & $4.8026(4)$ & $39.5(20)$ & $0.199(4)$\\
\cite{Bildsten1997} & $2.087113936(7)$$^{b}$ & $4.8$ & $39.6612(9)$ & -\\
\enddata
\tablecomments{$^{(a)}$ $\Phi_{\rm start}$ corresponds to the orbital phase at the start time of the observation. $^{(b)}$ At the time of $50 506.788423(7)$ MJD.}
\end{deluxetable*}

%% file: section3.tex
\section{Spectral variability along orbital phase} \label{sec:orbital_phase_resolved}

In order to investigate the orbital-phase variability of the spectrum, we divided the observation data into multiple pieces. Each piece has a duration of $500\;{\rm s}$, and 78 data sets were generated from $38.7\;{\rm ks}$ data. The spectrum extracted from each data set can be attributed to the mid-time of each interval.
Therefore, $\Phi_i$ ($1\leq i\leq78$) can be calculated by
\begin{eqnarray}
\Phi_i=\Phi_{\rm start}+\frac{500\;{\rm s}}{P_{\rm orb}}\left(i-0.5\right).
\label{eq:definition_of_orbital_phase}
\end{eqnarray}
In the calculation of the orbital phase, we used values listed in Table \ref{tab:orbital_parameters} for $\Phi_{\rm start}$ and $P_{\rm orb}$.
In the following analysis, 18 out of 78 orbital intervals were excluded because of very short exposures of $<30\;{\rm s}$, and the orbital-phase variability is investigated with the remaining 60 orbital intervals.

\subsection{Energy-resolved light curve}\label{sec:energy_resolved_light_curve}

\begin{figure*}[htb!]
\centering
\includegraphics[width=\linewidth]{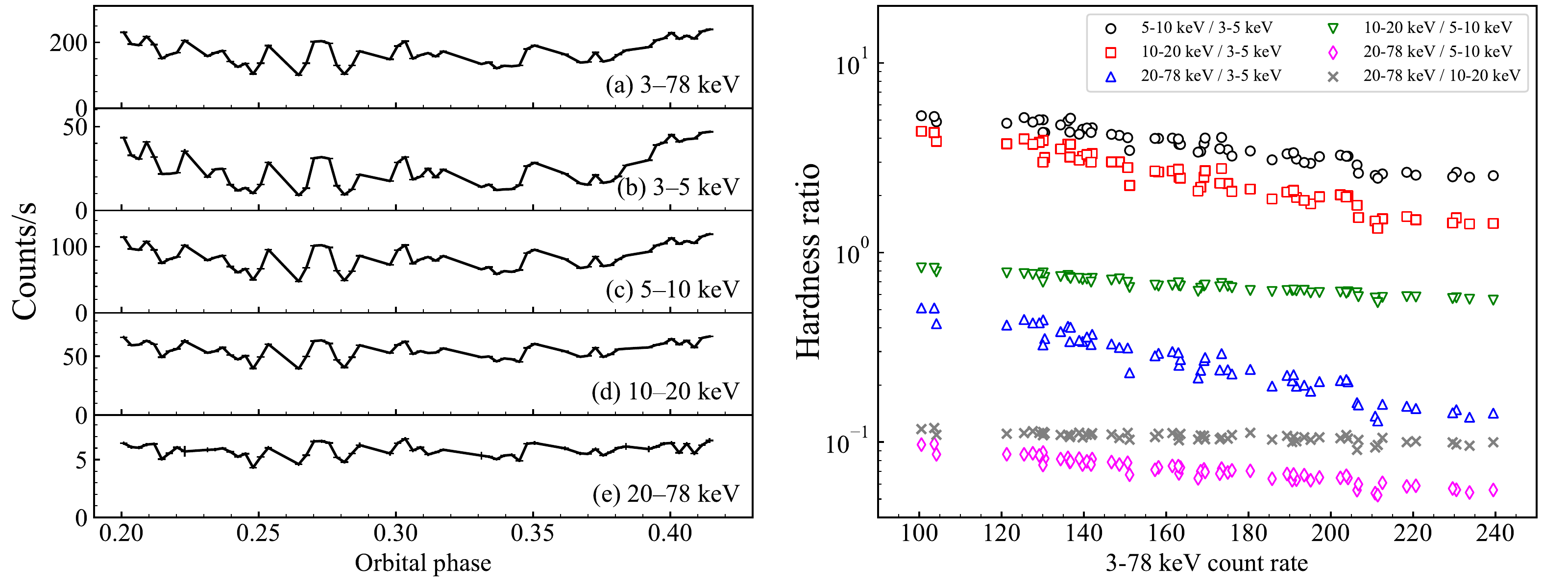}
\caption{(left) Energy-resolved light curves of (a) 3--78 keV, (b) 3--5 keV, (c) 5--10 keV, (d) 10--20 keV, and (e) 20--78 keV. Each point corresponds to a $500\;{\rm s}$ orbital interval defined in equation (\ref{eq:definition_of_orbital_phase}). FPMA and FPMB counts are added. (right) Hardness ratios as a function of 3--78 keV count rate, calculated from energy-resolved count rates at each orbital phase. Hardness ratios among multiple energy bands are plotted by different colors.}
\label{fig:lc_and_hr}
\end{figure*}

In order to examine the energy dependence of the orbital-phase variability, we first generated energy-resolved light curves based on the orbital intervals ($\Phi_1$--$\Phi_{78}$). The left panel of Figure \ref{fig:lc_and_hr} presents variations of count rate in 3--5, 5--10, 10--20, 20--78 keV, as well as the summation (3--78 keV) of them. The typical variation degree of a light curve can be evaluated by the ratio of the standard deviation to the average count rate.
The ratio (standard deviation / average count rate) was $0.43$ ($10.5/24.5$), $0.21$ ($17.7/83.6$), $0.12$ ($6.62/54.8$), and $0.09$ ($0.54/5.80$) for 3--5, 5--10, 10--20, and 20--78 keV, respectively.
The variation degree shows a monotonic decrease with the energy, which means the count variation along the orbital phase is more influenced by low-energy photons.

The hardness ratio is another parameter used to search for spectral variability.
It is used to express the relation between two specific energy bands, simply calculated by dividing the count rate of the higher energy band by that of the lower energy band.
The right panel of Figure \ref{fig:lc_and_hr} shows hardness ratios among different energy bands as a function of the total (3--78 keV) count rate.
While the 20--78 keV / 10--20 keV hardness ratio is almost constant, the other hardness ratios have clear negative correlations with the total count rate.
The nearly monotonic decreases of hardness ratios with the count rate indicate that the light curves of different energy bands are highly synchronized, with low energy bands more variable than high energy bands.


\subsection{Variability of spectral parameters}\label{sec:orbital_spectral_parameters}

In order to track the spectral variability more quantitatively, one needs to analyze the orbital-phase-resolved spectra.
In this section, we present a phenomenological spectral model that can be applied to all of the resolved spectra, and investigate the orbital-phase variation of spectral parameters.
We performed spectral fitting on the 60 spectra, extracted from the valid orbital intervals.
All of the spectra are re-binned to have at least 50 counts in each bin, and chi-squared statistics is adopted for spectral fitting.
FPMA and FPMB spectra are analyzed simultaneously by joint fitting.

The X-ray spectrum of Cen~X-3 can be roughly expressed by an absorbed power law with a high-energy cut-off around $\sim 15\;{\rm keV}$, accompanied by a soft excess, three Fe emission lines in 6--7 keV, and the cyclotron resonance scattering feature (CRSF) at $\sim 30\;{\rm keV}$ \citep{Burderi2000, Naik2011, Tomar2021}.
We employed Fermi-Dirac cut-off power law \citep{Tanaka1986} to fit the continuum, which is expressed by
\begin{eqnarray}
f_{\rm FDcut}(E)=\frac{E^{-\Gamma}}{\exp\left(\frac{E-E_{\rm c}}{E_{\rm f}}\right)+1},
\label{eq:fermi_dirac}
\end{eqnarray}
where $\Gamma$, $E_{\rm c}$, and $E_{\rm f}$ are the photon index, the cut-off energy, and the folding energy, respectively.
The photo-electric absorption dominant in the low energy band is reproduced by {\tt phabs} model in {\tt XSPEC}, expressed by
\begin{eqnarray}
f_{\rm phabs}(E)=\exp\left[-\left(\nhim+\nh\right)\sigma(E)\right],
\label{eq:phabs}
\end{eqnarray}
where $\nhim$, $\nh$, and $\sigma(E)$ are the hydrogen column density of the interstellar medium, that of the stellar wind, and the photo-electric cross section based on \cite{Verner1996}, respectively.
For other additional components, we applied simplification to the previously suggested model because each spectrum has lower statistics than the phase-averaged spectrum.
The soft excess and CRSF were not included in the model due to low statistics.
The three Fe emission lines are simplified to a single Gaussian model, which is defined by
\begin{eqnarray}
f_{\rm gauss}(E)=\frac{I_{\rm Fe}}{\sqrt{2\pi}\sigma_{\rm Fe}}\exp\left[\frac{\left(E-E_{\rm Fe}\right)^2}{2\sigma_{\rm Fe}^2}\right],
\label{eq:gauss}
\end{eqnarray}
where $I_{\rm Fe}$, $E_{\rm Fe}$, and $\sigma_{\rm Fe}$ are the normalization factor, the central energy, and the standard deviation of the Fe line, respectively.
Therefore, the spectral model we employed is
\begin{eqnarray}
S(E)\propto f_{\rm phabs}\times\left[f_{\rm FDcut}+f_{\rm gauss}\right].
\label{eq:model_expression}
\end{eqnarray}

We performed spectral fitting on all of the 60 orbital-phase-resolved spectra, using equation (\ref{eq:model_expression}). All of the spectral parameters are set free except that $\nhim=1.1\times10^{22}\;{\rm cm^{-2}}$, $\sigma_{\rm Fe}=0$, and $E_{\rm Fe}=6.4\;{\rm keV}$. The first fixed parameter comes from the galactic hydrogen column density towards the source \citep{HI4PI2016}. The second one is based on the assumption that the line width is small enough compared with the energy resolution of \nustar. The last one comes from the fact that the neutral Fe K-emission line is dominant compared to the H-like or He-like Fe K-emission lines \citep{Naik2011}.
The cross-normalization factor between FPMA and FPMB is treated as another fitting parameter.
In the analysis, we performed fitting on 4--78 keV and excluded 3--4 keV photons since this band can be strongly affected by the soft excess component.
It is difficult to evaluate the contribution of the soft excess with \nustar\ alone because its origin is a blackbody with a temperature of $\sim0.1\;{\rm keV}$ \citep{Burderi2000}.

\begin{figure}[htb!]
\centering
\includegraphics[width=240pt]{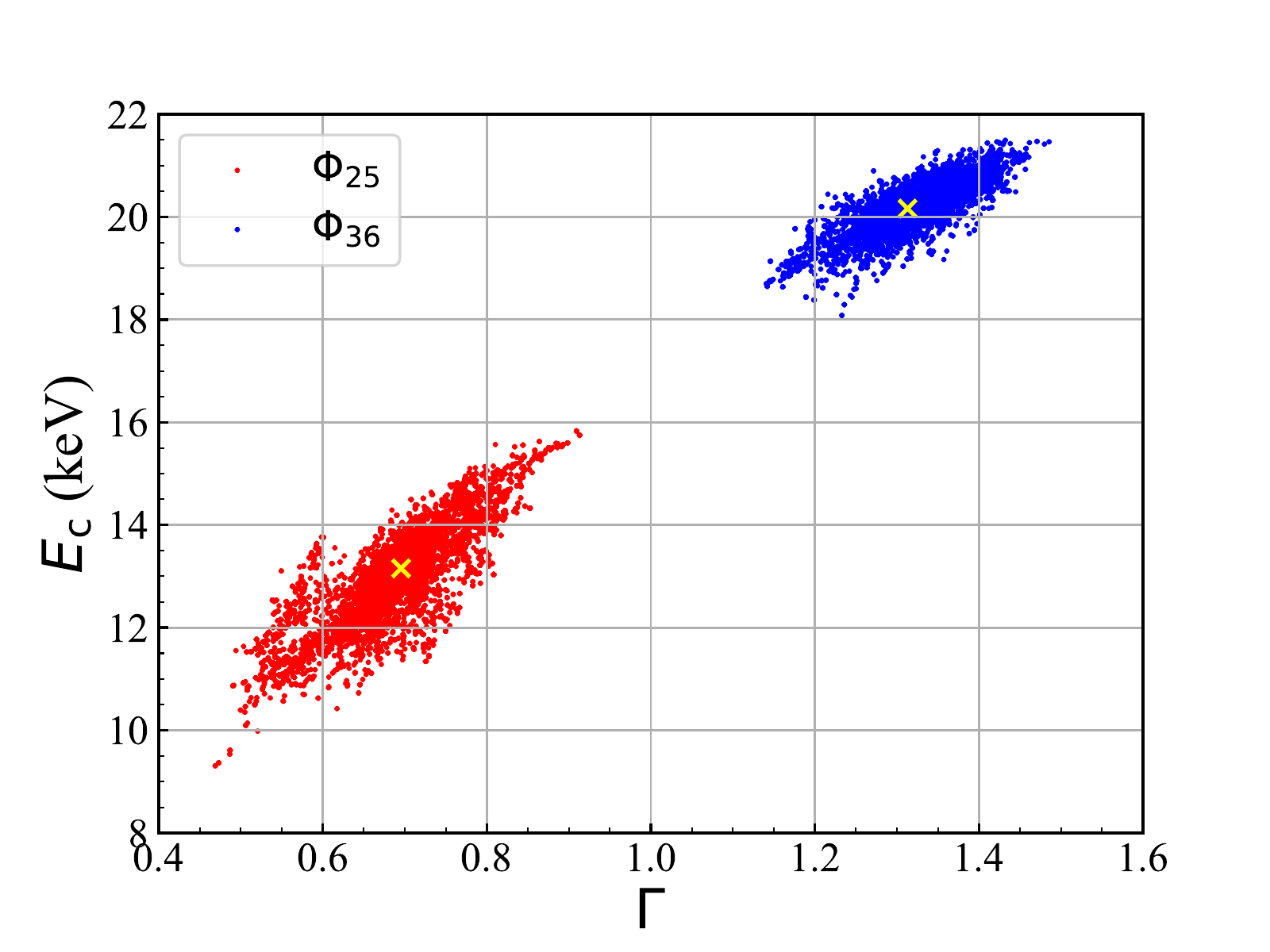}
\caption{Parameter correlation of photon index ($\Gamma$) and cut-off energy ($E_{\rm c}$) calculated from 10000 step MCMC. 1000 steps are discarded beforehand as an initial burn-in phase. Red and blue points represent orbital phases with the hardest ($\Phi_{25}$) and softest photon index ($\Phi_{36}$), respectively.}
\label{fig:mcmc_PI_Ec}
\end{figure}

For the photon index, two ways of analysis were carried out.
We treated $\Gamma$ as a free parameter in one analysis and fixed it in the other.
The fixed photon-index analysis was carried out because the strong parameter correlation between $\Gamma$ and $E_{\rm c}$ may prevent us from yielding correct best-fit parameters.
The strong correlation between these two parameters reproduced by MCMC (Markov Chain Monte Carlo) method is presented in Figure \ref{fig:mcmc_PI_Ec}, in which we picked up two characteristic orbital intervals, that with the hardest photon index ($\Phi_{25}$) and softest photon index ($\Phi_{36}$).
In the fixed photon-index analysis, we fixed the photon index to $\Gamma=1.21$, which is the best-fit value of the phase-averaged spectrum derived by \cite{Tomar2021}.

\begin{figure}[htb!]
\centering
\includegraphics[width=240pt]{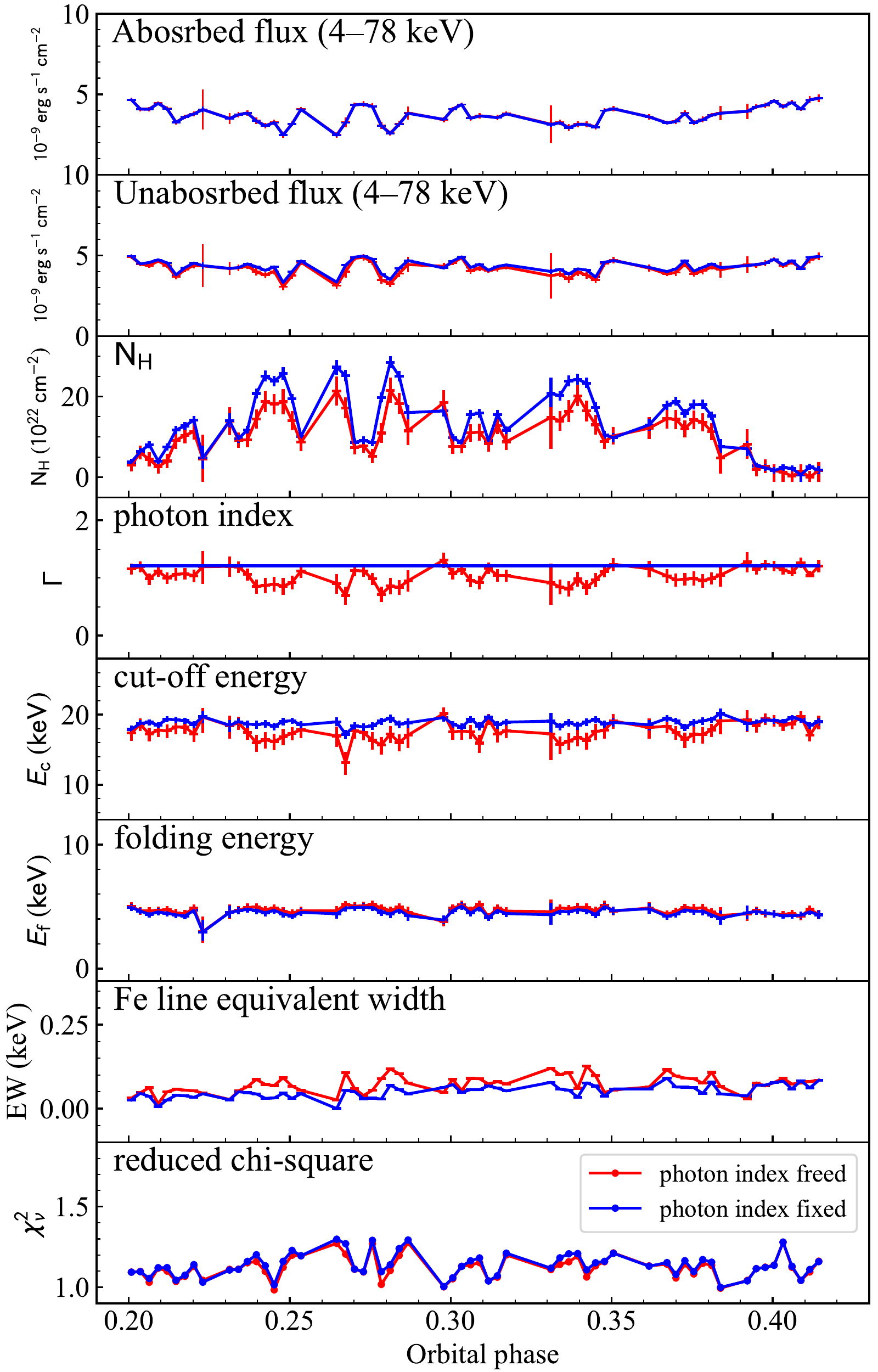}
\caption{Orbital-phase variation of spectral parameters derived from 4--78 keV orbital-phase-resolved spectra. The assumed model is Fermi-Dirac cut-off power-low with an Fe emission line (see text). Best-fit parameters with freed photon index (red) and fixed photon index of $\Gamma=1.21$ (blue) are plotted together. Error bars of fitting parameters represent 90\% confidence levels.}
\label{fig:orbital_spectral_parameters}
\end{figure}

In Figure \ref{fig:orbital_spectral_parameters}, we show the orbital-phase variability of spectral parameters derived from 4--78 keV spectral fitting.
We represent the free photon-index analysis with red lines and the fixed photon-index analysis with blue lines.
The fitting results of the former analysis returned acceptable reduced chi-squares with an average value of $\chi_{\nu}^2=1.12$.
In the latter case, the reduced chi-squares got slightly worse, with an acceptable average value of $\chi_{\nu}^2=1.14$.
As a result of fixing $\Gamma$, other parameters such as $E_{\rm c}$, $E_{\rm f}$, and the equivalent width of the Fe line got much more robust to the orbital phase.
Here, we do not rule out either of the assumptions and denote the former as Case 1 (free photon index) and the latter as Case 2 (fixed photon index).

\begin{figure*}[htb!]
\centering
\includegraphics[width=\linewidth]{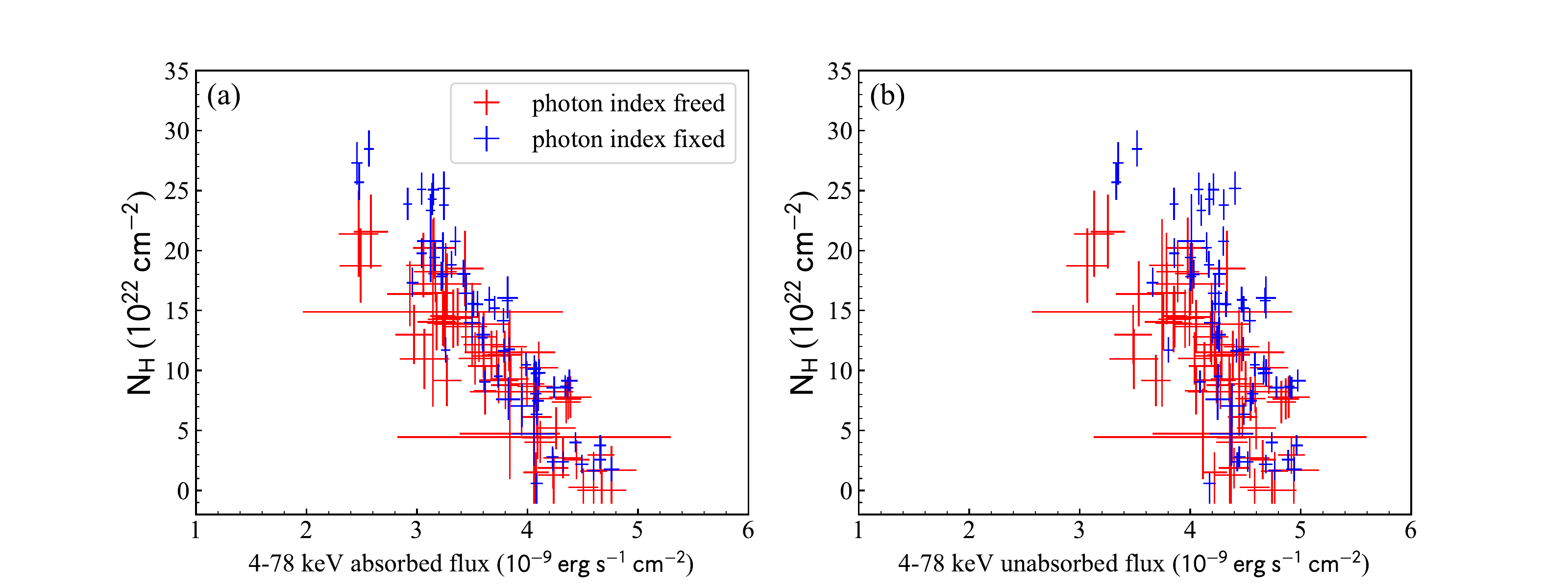}
\caption{Relation between (a) 4--78 keV absorbed flux and $\nh$, (b) 4--78 keV unabsorbed flux and $\nh$, derived from orbital-phase-resolved spectroscopy. Best-fit parameters with freed photon index (red) and fixed photon index of $\Gamma=1.21$ (blue) are plotted together. Error bars of fitting parameters represent 90\% confidence levels.}
\label{fig:flux_to_nh}
\end{figure*}

Whether freeing or fixing the photon index, $\nh$ shows the largest variability among the spectral parameters.
It varies from $0$ to $2.2\times10^{23}\;{\rm cm^{-2}}$ for Case 1 and from $0$ to $2.8\times10^{23}\;{\rm cm^{-2}}$ for Case 2.
For both cases, the variation of $\nh$ is clearly correlated to the variability of the absorbed flux.
As presented in Figure \ref{fig:flux_to_nh}(a), these two parameters are negatively correlated throughout multiple orbital phases, which suggests that the flux variation is mainly caused by different degrees of absorption.
On the other hand, the unabsorbed flux has a weak correlation with $\nh$, as shown in Figure \ref{fig:flux_to_nh}(b).
This result is consistent with the analysis of energy-resolved light curves in Section \ref{sec:energy_resolved_light_curve} because the difference in absorption degree more likely influences low-energy photons.

From the variability of the absorbed and unabsorbed flux, we can estimate the effect of stellar wind on the flux variability.
Here, we evaluate the typical flux variation degree by the standard deviation divided by the average flux, which is calculated from the 60 orbital intervals.
The variation degrees of the absorbed flux were 14.9\% for both of the analysis cases.
The variation degree of the unabsorbed flux was 10.3\% and 8.6\% for Case 1 and 2, respectively, which can be regarded as intrinsic variability of the source.
Assuming the intrinsic and extrinsic variabilities are independent, the flux variation caused by stellar wind absorption is 10.8\% and 12.2\% for Case 1 and 2, respectively.
Therefore, for the NuSTAR energy band, the effect of stellar wind on the flux variability is comparable to the intrinsic variability.



%% file: section4.tex
\section{Spectral variability along spin phase} \label{sec:spin_phase_resolved}

In addition to the orbital-phase variability, we inspected the spin-phase variability.
In order to track this short-term variability, we focused on both energy-resolved pulse profiles and spin-phase-resolved spectra.
In this paper, we denote the spin phase by $\phi$ $(0.0\leq\phi<1.0)$ to distinguish it from the orbital phase $\Phi$.

\subsection{Energy-resolved pulse profile}\label{sec:pulse_profile}

\begin{figure*}[htb!]
\centering
\includegraphics[width=\linewidth]{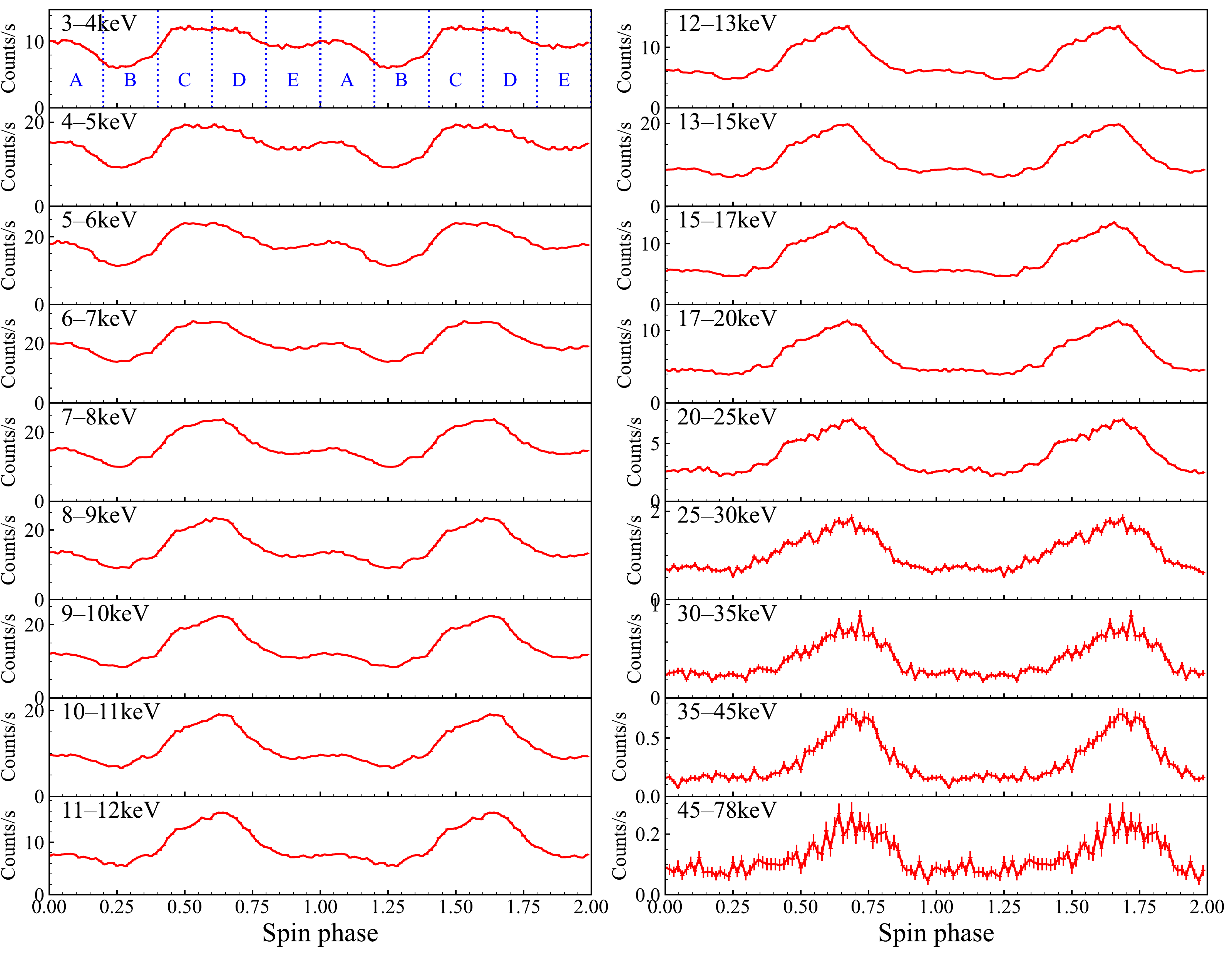}
\caption{Energy-resolved pulse profiles generated with binary parameters listed in Table \ref{tab:orbital_parameters}. The summation of FPMA and FPMB counts are plotted. The spin phase A--E defined in Section \ref{sec:spin_ratio_spectrum} are also presented.}
\label{fig:pulse_profiles}
\end{figure*}

The energy dependence of the spin-phase variability can be examined by energy-resolved pulse profiles.
We divided 3--78 keV into 16 bands and generated pulse profiles by folding each energy-resolved light curve by the spin period of $P_{\rm spin}=4.8026\;{\rm s}$ (see Table \ref{tab:orbital_parameters}).
The zero point of the spin phase was set so that $\phi=0.0$ at $t_{\rm binary}=57356.0\;{\rm MJD}$.
Figure \ref{fig:pulse_profiles} presents the energy-resolved pulse profiles of the 18 bands. All the energy bands clearly show the pulsations, and the shapes of pulse profiles gradually change as energy increases. The profiles appear to show moderate pulsations and double-peaked shapes in low-energy bands. As energy goes up, the pulsation becomes sharp, and the second peak declines, gradually shifting to a single-peaked shape, as already reported by \cite{Tomar2021}.

In order to quantify these trends, we define the rms pulse fraction by
\begin{eqnarray}
{\rm PF_{rms}}=\frac{2\sqrt{\sum_{k=1}^{k_{\rm max}}\left(\left(a_k^2+b_k^2\right)-\left(\sigma_{a_k}^2+\sigma_{b_k}^2\right)\right)}}{a_0},
\label{eq:rms_pulse_fraction}
\end{eqnarray}
where $a_k$ and $b_k$ are Fourier coefficients written by
\begin{eqnarray}
a_k&=&\frac{1}{N}\sum_{j=1}^{N}p_j\cos\left(\frac{2\pi kj}{N}\right)\nonumber\\
b_k&=&\frac{1}{N}\sum_{j=1}^{N}p_j\sin\left(\frac{2\pi kj}{N}\right),
\label{eq:fourier_coeffcients}
\end{eqnarray}
and $\sigma_{a_k}^2$ and $\sigma_{b_k}^2$ are defined by
\begin{eqnarray}
\sigma_{a_k}^2&=&\frac{1}{N^2}\sum_{j=1}^{N}\sigma_{p_j}^2\cos^2\left(\frac{2\pi kj}{N}\right)\nonumber\\
\sigma_{b_k}^2&=&\frac{1}{N^2}\sum_{j=1}^{N}\sigma_{p_j}^2\sin^2\left(\frac{2\pi kj}{N}\right).
\end{eqnarray}
$N$, $p_j$, and $\sigma_{p_j}$ are the number of bins, the photon count rate at $j$-th bin, and the Poisson variance at $j$-th bin, respectively. The derivation of these equations is explained in \cite{Tendulkar2015} and Appendix 1 of \cite{An2015}. The contribution of $k$-th Fourier coefficient number can be calculated by
\begin{eqnarray}
c_k=\sqrt{a_k^2+b_k^2}.
\label{eq:fourier_coeffcients_sum}
\end{eqnarray}
The definition of ${\rm PF_{rms}}$ employs the correction term described by $\sigma_{a_k}$ and $\sigma_{b_k}$, which leads to a robust and accurate evaluation of the pulse fraction with a noisy data \citep[see][]{An2015}.

\begin{figure*}[htb!]
\centering
\includegraphics[width=\linewidth]{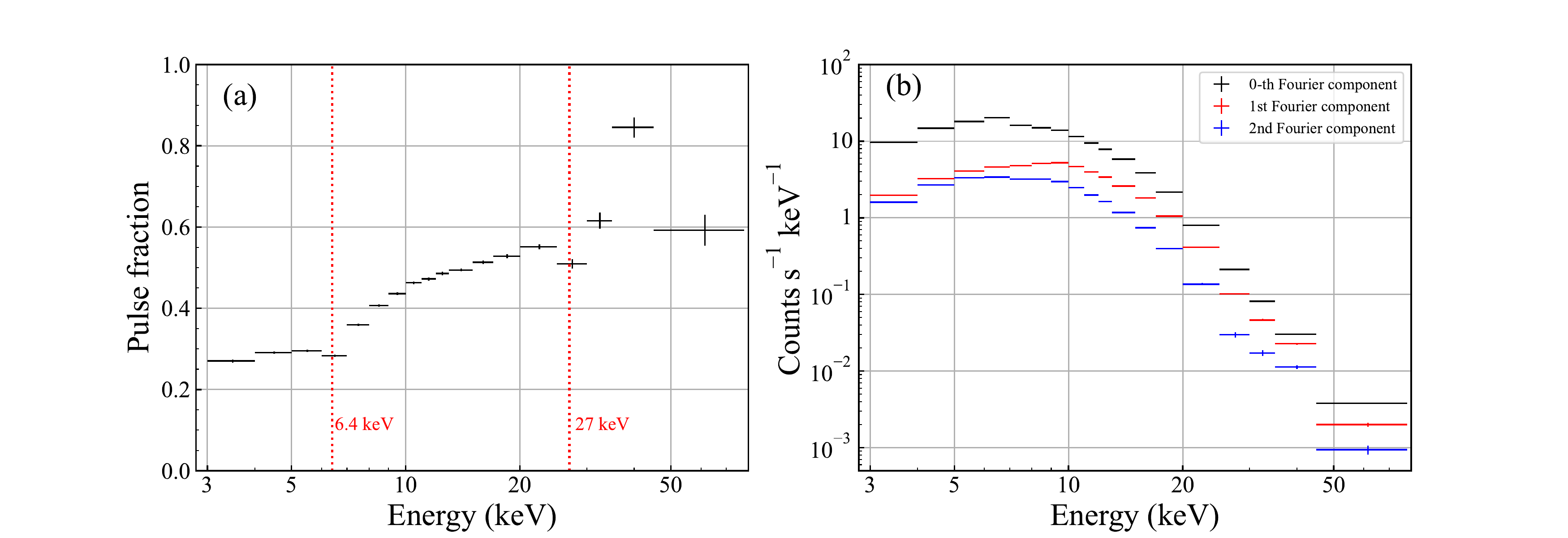}
\caption{(a) Pulse fraction ${\rm PF_{rms}}$ as a function of energy, calculated by equation (\ref{eq:rms_pulse_fraction}). The energies corresponding to the Fe emission line and CRSF are presented by red dotted lines. (b) Contributions of the 1st (red) and 2nd (blue) Fourier components compared to the phase-averaged (0-th, black) spectrum. The definition of Fourier coefficients is explained in equation (\ref{eq:fourier_coeffcients}) and (\ref{eq:fourier_coeffcients_sum}).}
\label{fig:pulse_profile_analysis}
\end{figure*}

Figure \ref{fig:pulse_profile_analysis} presents the results of the detailed Fourier analysis applied to the energy-resolved pulse profiles.
Figure \ref{fig:pulse_profile_analysis}(a) shows the energy dependence of the pulse fraction, which is defined by equation (\ref{eq:rms_pulse_fraction}).
In the range of 3--45 keV, the pulse fraction monotonically increases as energy goes up with the exceptions of two characteristic bands, 6--7 keV and 25--30 keV.
These bands correspond to the Fe K-emission line ($\sim6.4\;{\rm keV}$) and the CRSF ($\sim27\;{\rm keV}$), respectively.
Figure \ref{fig:pulse_profile_analysis}(b) shows the contributions of the 1st and 2nd Fourier components compared to the 0-th component (the phase-averaged spectrum).
The 1st and 2nd Fourier coefficients are comparable at the lowest energy band, while the 1st Fourier component becomes dominant as energy increases.
This result is consistent with the single-peaked profile at higher energy bands, as already seen in Figure \ref{fig:pulse_profiles}.

\subsection{Ratio spectrum} \label{sec:spin_ratio_spectrum}

As an indicator to track the variability of the spectrum, we define ``ratio spectrum'', which is a count ratio profile of one spectrum with respect to a reference spectrum.
This barometer is employed in \cite{Kondo2021}, which analyzed spin-phase-resolved spectra of Her~X-1.
Here, we define $S(E;\;\phi_{1},\;\phi_{2})$ as the count spectrum per area of a spin-phase interval ranging from $\phi_1$ to $\phi_2$.
We also denote $S_{\rm ave}(E)$ as the phase-averaged spectrum per area.
The ratio spectrum of a specific spin-phase interval ($\phi=\phi_1-\phi_2$) can be expressed by
\begin{eqnarray}
R(E;\;\phi_1,\;\phi_2)=\frac{S(E;\;\phi_1,\;\phi_2)}{S_{\rm ave}(E)}.
\end{eqnarray}
This barometer can examine how the spin-phase variability differs depending on the energy.
If the source emission were completely stable, the ratio spectrum would equal unity, namely, $R(E;\;\phi_1,\;\phi_2)=1$ at every energy. The calculation of the ratio spectrum is applied to the summed spectrum of two focal plane detectors, FPMA and FPMB.

\begin{figure*}[htb!]
\centering
\includegraphics[width=\linewidth]{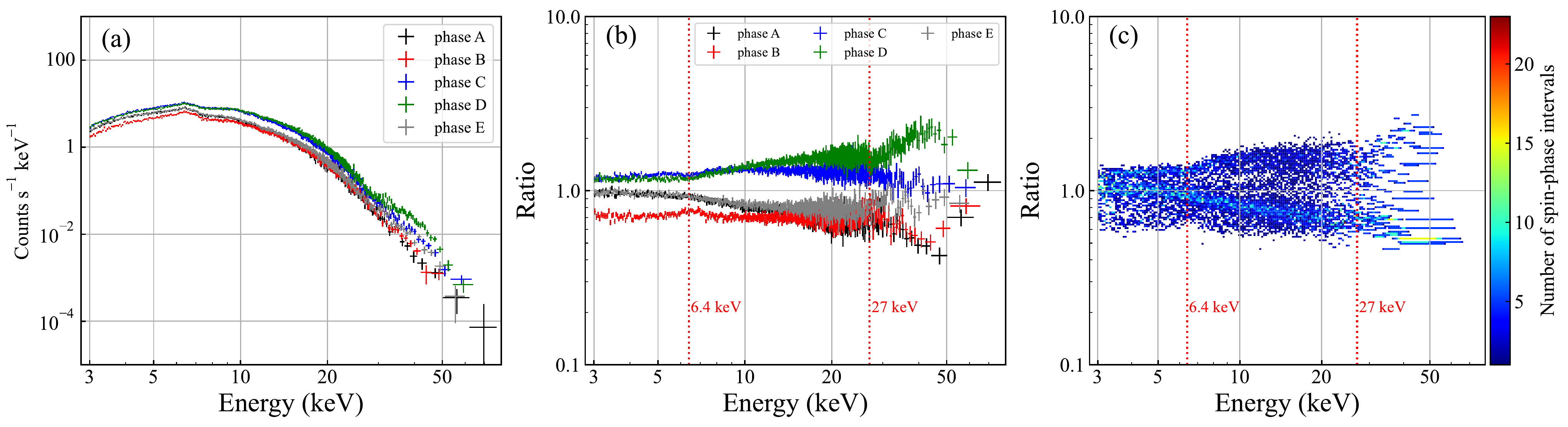}
\caption{(a) Spin-phase-resolved FPMA spectra of phase A (black), B (red), C (blue), D (green), and E (grey). (b) Ratio spectra for phase A--E. FPMA and FPMB spectra are added before the calculation. (c) Scatter plot of ratio spectra generated from 20 sets of spin-phase-resolved spectrum, each of which covers a spin phase interval of $\Delta\phi=0.05$. The color bar represents the number of ratio spectra that have a specific value at each energy. The energies corresponding to the Fe emission line and CRSF are presented by red dotted lines.}
\label{fig:spin_ratio_spectrum}
\end{figure*}

We divided the whole spin phase into five intervals to analyze the spin-phase variability.
Hereafter, we denote the five spin-phase intervals as phase A ($\phi=0.0$--$0.2$), B ($\phi=0.2$--$0.4$), C ($\phi=0.4$--$0.6$), D ($\phi=0.6$--$0.8$), and E ($\phi=0.8$--$1.0$).
The definitions of spin phase intervals A--E are also presented in the 3--4 keV pulse profile in Figure \ref{fig:pulse_profiles}.
From the pulse profiles presented in Figure \ref{fig:pulse_profiles}, phase D corresponds to the pulse maximum and B to the pulse minimum.
We generated spin-phase-resolved spectra of spin-phase intervals A--E and calculated ratio spectra for each spin-phase interval.
Figure \ref{fig:spin_ratio_spectrum}(a) shows the five spin-phase-resolved spectra, and Figure \ref{fig:spin_ratio_spectrum}(b) presents the ratio spectra of the five spin-phase intervals.
The ratio spectra show the presence of large variations in high energy bands, suggesting high-flux phases have harder spectra than low-flux phases.
There are also characteristic features around the Fe line ($\sim6.4\;{\rm keV}$) and CRSF ($\sim27\;{\rm keV}$) band, where the ratio gets closer to unity than nearby bands.

We also generated 20 ratio spectra from 20 spin-phase-resolved spectra, each covering a spin-phase interval of $\Delta\phi=0.05$.
Figure \ref{fig:spin_ratio_spectrum}(c) presents the scatter plot of the 20 ratio spectra.
It shows more significant spectral variability in higher energy bands and dented features around the Fe line and CRSF band.
Therefore, those properties seen in the five ratio spectra (Figure \ref{fig:spin_ratio_spectrum}b) still hold even when the division of the spin phase becomes finer.
The results derived from ratio spectra are consistent with that of energy-resolved pulse profiles (Section \ref{sec:pulse_profile}).
This is because the larger variation at higher energy and dented features around the Fe line and CRSF band correspond to the monotonic increase of the pulse fraction and low pulse fraction at these two bands, respectively (Figure \ref{fig:pulse_profile_analysis}a).

\subsection{Variability of spectral parameters}\label{sec:spin_spectral_parameters}

The qualitative features of the spin-phase spectral variability have been investigated through energy-resolved pulse profiles and ratio spectra (Sections \ref{sec:pulse_profile} and \ref{sec:spin_ratio_spectrum}).
They have revealed that the source spectrum has a harder photon index at higher flux phases and softer at lower flux.
It has also become clear that the spin-phase variations of the Fe emission line and CRSF are moderate compared to nearby bands, which means lower equivalent width of Fe line and deeper absorption of CRSF at high-flux phases.
In order to examine these properties more quantitatively, we performed spectral fitting to spin-phase-resolved spectra.
We picked up spectra of the spin phase A--E (see the definition in Section \ref{sec:spin_ratio_spectrum}) and tracked the variations of spectral parameters along the spin phase.

The assumed spectral model here is basically the same as that applied to the orbital-phase-resolved analysis (equation \ref{eq:model_expression}).
The difference is the inclusion of the Gaussian absorption line model ({\tt gabs} in {\tt XSPEC}) to describe the CRSF.
The variability of the CRSF can be evaluated because of the relatively rich statistics compared to the orbital-phase-resolved analysis.
Therefore, the model adopted here is
\begin{eqnarray}
S(E)\propto f_{\rm phabs}\times\left(f_{\rm FDcut}+f_{\rm gauss}\right)\times f_{\rm gabs},
\label{eq:spin_phase_model}
\end{eqnarray}
where $f_{\rm gabs}$ denotes the absorption line feature and is expressed by
\begin{eqnarray}
f_{\rm gabs}(E)=\exp\left\{-\tau_{\rm cyc}\exp\left[-\frac{\left(E-E_{\rm cyc}\right)^2}{2\sigma_{\rm cyc}^2}\right]\right\}
\label{eq:gabs}
\end{eqnarray}
$f_{\rm FDcut}$, $f_{\rm phabs}$, and $f_{\rm gauss}$ are defined in equations (\ref{eq:fermi_dirac}), (\ref{eq:phabs}) and (\ref{eq:gauss}), respectively.
$\tau_{\rm cyc}$, $E_{\rm cyc}$, and $\sigma_{\rm cyc}$ represent the optical depth at the line center, the central energy, and the standard deviation of the CRSF, respectively.
All of the parameters were set free except that $\nhim=1.1\times10^{22}\;{\rm cm^{-2}}$ \citep{HI4PI2016} and $\nh=3.74\times10^{22}\;{\rm cm^{-2}}$.
The latter constraint comes from the average $\nh$ calculated from the orbital-phase-resolved spectroscopy with a fixed photon index (Case 2 in Section \ref{sec:orbital_spectral_parameters}).
We assume that $\nh$ originates from the absorption by the stellar wind and is independent of the spin phase.
$\sigma_{\rm Fe}$ and $E_{\rm Fe}$ are set free because of sufficient photon counts, which were frozen in the orbital-phase-resolved spectroscopy. FPMA and FPMB spectra are analyzed simultaneously by joint fitting, setting a cross-normalization factor as another fitting parameter.

\begin{figure*}
\includegraphics[width=\linewidth]{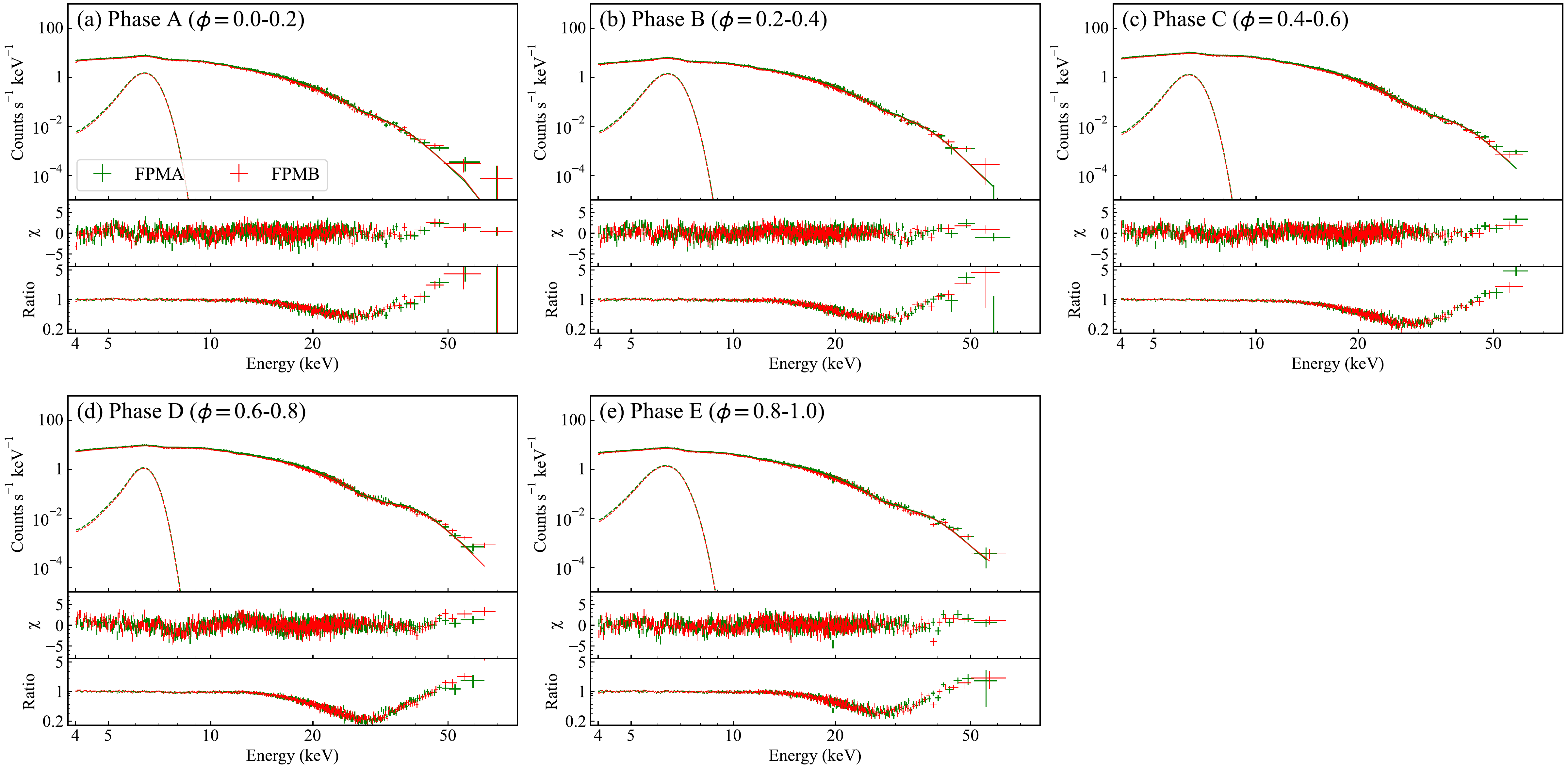}
\caption{4--78 keV fitting results of spin-phase-resolved spectra of (a) phase A, (b) phase B, (c) phase C, (d) phase D, and (e) phase E. For the model expression, see equation (\ref{eq:spin_phase_model}). The green and red plots represent FPMA and FPMB data, respectively. The contributions of the Gaussian component are plotted by dashed lines. The middle panels show residuals between data and the best-fit model, and the bottom panels show ratios between data and model when the the contribution of  CRSF is completely excluded from the model.}
\label{fig:spin_phase_fitting}
\end{figure*}

\begin{figure*}
\includegraphics[width=\linewidth]{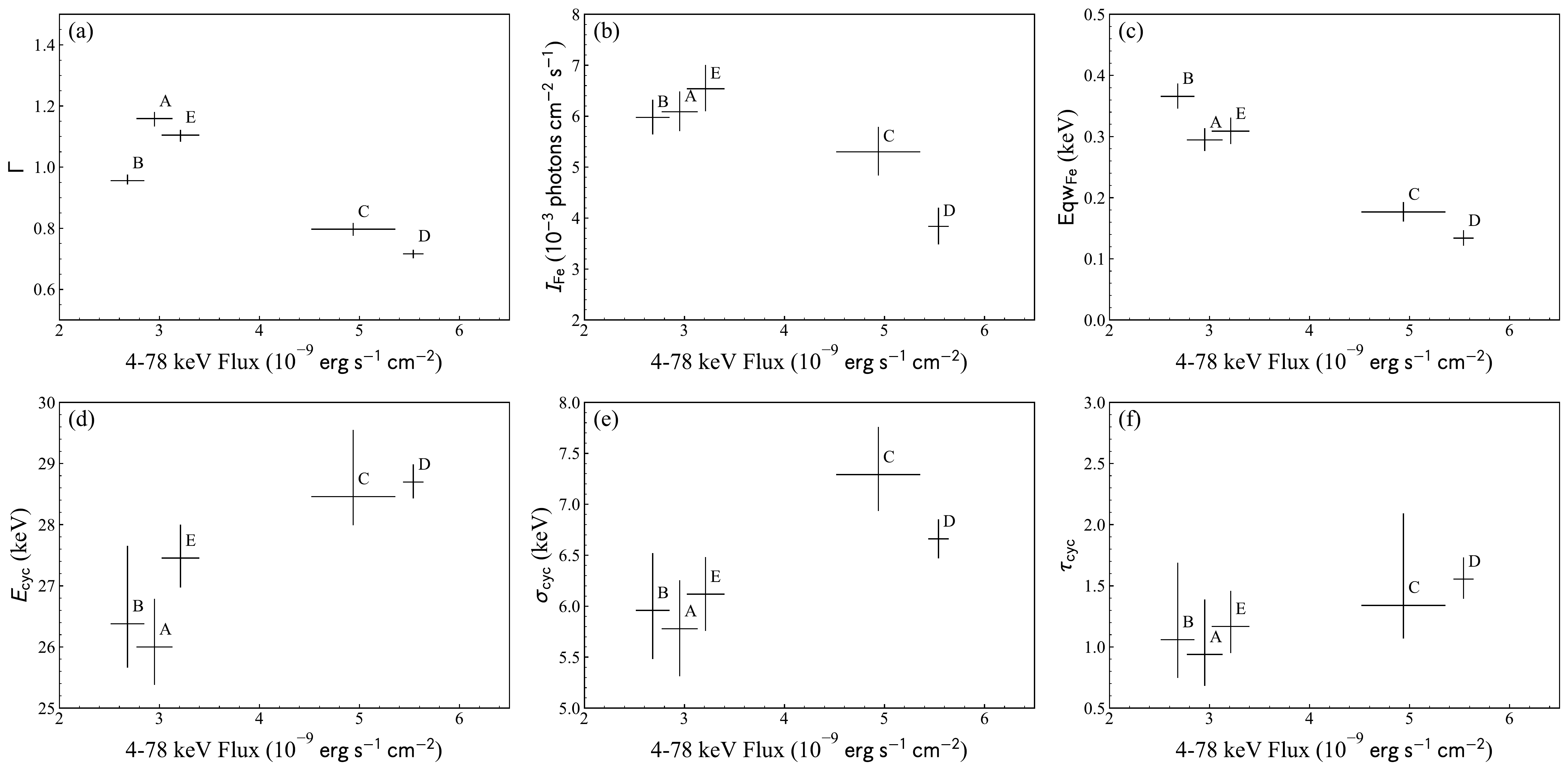}
\caption{Spectral parameter variations of spin-phase-resolved spectroscopy plotted as a function of the source flux, derived from spin phase A--E (see text). The presented spectral parameters are (a) photon index, (b) normalization of Fe line, (c) equivalent width of Fe line, (d) central energy of CRSF, (e) standard deviation of CRSF, and (f) optical depth of CRSF. Error bars denote 90\% confidence levels.}
\label{fig:spin_phase_parameter_correlation}
\end{figure*}

Figure \ref{fig:spin_phase_fitting} shows the fitting results of the five spectra, which correspond to the spin phase A--E, respectively. The residuals have no distinct features, and the absorption line features caused by the CRSF are well reproduced. The best-fit parameters are presented in Table \ref{tab:spin_phase_fitting_parameters}. All the five spectra returned acceptable $\chi_\nu^2$ of $1.09$--$1.30$.
In Table \ref{tab:spin_phase_fitting_parameters}, we can see some variations of spectral parameters depending on the spin phase, especially differences between the pulse maximum (phase D) and the pulse minimum (phase B).
We picked up some spectral parameters and presented their correlations with the source flux in Figure \ref{fig:spin_phase_parameter_correlation}.

$\Gamma$ is an important parameter that measures the spectral hardening due to Comptonization inside the accretion column.
According to Figure \ref{fig:spin_phase_parameter_correlation}(a), it varies from $0.72$ to $1.16$ and is negatively correlated to the continuum flux.
The harder spectra at the high-flux phases (phase C, D) and softer ones at the low-flux phases (phase A, B, E) are consistent with the analysis results of pulse profiles (Section \ref{sec:pulse_profile}) and ratio spectra (Section \ref{sec:spin_ratio_spectrum}), both of which suggested larger variability in higher energy bands.
Because the cut-off and folding energy do not show any significant variation, the photon index variation originates from the difference in the optical depth of Comptonization, and the effective electron temperature does not change.

The Fe line shows a distinct variability along the spin phase.
As is shown in Figure \ref{fig:spin_phase_parameter_correlation}(b) and (c), both the intensity and the equivalent width of the Fe line are negatively correlated with the source flux.
This feature is also seen in the pulse profiles and ratio spectra as low variability around the Fe line band.
At the pulse maximum, the intrinsic strength and the equivalent width of the Fe line were $\sim60\%$ and $\sim35\%$ of those at the pulse minimum, respectively.

The CRSF also presents different contributions at different spin phase, as presented in Figure \ref{fig:spin_phase_parameter_correlation}(d), (e), and (f).
The high-flux phases (C, D) have significantly higher $E_{\rm cyc}$ and larger $\sigma_{\rm cyc}$ than the low-flux phases (A, B, E).
The variation degrees of $E_{\rm cyc}$ and $\sigma_{\rm cyc}$ are $\sim10\%$ and $\sim25\%$, respectively.
The variation of the optical depth $\tau_{\rm cyc}$, on the other hand, is not significant due to large uncertainties, but its positive correlation with the source flux is strongly suggested by low variability at CRSF band derived by energy-resolved pulse fraction (Figure \ref{fig:pulse_profile_analysis}a) and ratio spectra (Figure \ref{fig:spin_ratio_spectrum}).

\begin{deluxetable*}{cccccc}
\tablecaption{Best-fit parameters derived from spin-phase-resolved spectroscopy.\label{tab:spin_phase_fitting_parameters}}
\tablewidth{0pt}
\tablehead{
\colhead{Parameter$^{a}$} & \colhead{\begin{tabular}{c}phase A\\$(\phi=0.0$--$0.2$)\end{tabular}} & \colhead{\begin{tabular}{c}phase B\\$(\phi=0.2$--$0.4$)\end{tabular}} & \colhead{\begin{tabular}{c}phase C\\$(\phi=0.4$--$0.6$)\end{tabular}} & \colhead{\begin{tabular}{c}phase D\\$(\phi=0.6$--$0.8$)\end{tabular}} & \colhead{\begin{tabular}{c}phase E\\$(\phi=0.8$--$1.0$)\end{tabular}}
}
\startdata
$\nhim$ ($10^{22}\;{\rm cm^{-2}}$) & \multicolumn{5}{c}{$1.1$ (fixed)} \\
$\nh$ ($10^{22}\;{\rm cm^{-2}}$) & \multicolumn{5}{c}{$3.74$ (fixed)} \\
$\Gamma$ & $1.16^{+0.02}_{-0.03}$ & $0.96^{+0.02}_{-0.01}$ & $0.80\pm0.02$ & $0.72\pm0.01$ & $1.10\pm0.02$\\
$E_{\rm c}$ (keV) & $24^{+3}_{-2}$ & $25^{+4}_{-2}$ & $24^{+5}_{-2}$ & $27\pm1$ & $26\pm2$\\
$E_{\rm f}$ (keV) & $4.4^{+0.4}_{-0.7}$ & $4.1^{+0.5}_{-0.8}$ & $4.9^{+0.4}_{-1.0}$ & $5.1^{+0.3}_{-0.3}$ & $5.0^{+0.4}_{-0.5}$\\
$E_{\rm Fe}$ (keV) & $6.39^{+0.02}_{-0.03}$ & $6.42\pm0.02$ & $6.35\pm0.04$ & $6.36\pm0.03$ & $6.33\pm0.03$\\
$\sigma_{\rm Fe}$ (keV) & $0.42\pm0.03$ & $0.44\pm0.03$ & $0.41^{+0.05}_{-0.04}$ & $0.32\pm0.04$ & $0.49\pm0.04$\\
$I_{\rm Fe}$ ($10^{-3}\;{\rm photons\;cm^{-2}\;s^{-1}}$) & $6.1\pm0.4$ & $6.0\pm0.3$ & $5.3\pm0.5$ & $3.8\pm0.4$ & $6.5^{+0.5}_{-0.4}$\\
${\rm Eqw}_{\rm Fe}^{b}$ (keV) & $0.29\pm0.02$ & $0.37^{+0.02}_{-0.02}$ & $0.18\pm0.02$ & $0.13\pm0.01$ & $0.31\pm0.02$\\
$E_{\rm cyc}$ (keV) & $26.0^{+0.8}_{-0.6}$ & $26.4^{+1.3}_{-0.7}$ & $28.5^{+1.1}_{-0.5}$ & $28.7\pm0.3$ & $27.5\pm0.5$\\
$\sigma_{\rm cyc}$ (keV) & $5.8\pm0.5$ & $6.0^{+0.6}_{-0.5}$ & $7.3^{+0.5}_{-0.4}$ & $6.7\pm0.2$ & $6.1\pm0.4$\\
$\tau_{\rm cyc}$ & $0.9^{+0.5}_{-0.3}$ & $1.1^{+0.6}_{-0.3}$ & $1.3^{+0.8}_{-0.3}$ & $1.6\pm0.2$ & $1.2^{+0.3}_{-0.2}$\\
Flux$^{c}$ ($10^{-9}\;{\rm erg\;s^{-1}\;cm^{-2}}$) & $3.0\pm0.2$ & $2.7\pm0.2$ & $4.9\pm0.3$ & $5.5\pm0.1$ & $3.2\pm0.2$\\
$\chi_{\nu}^2$ (d.o.f) & 1.15 (1000) & 1.07 (1001) & 1.09 (1117) & 1.30 (1173) & 1.16 (1033)\\
\enddata
\tablecomments{Errors denote 90\% confidence levels. $^{(a)}$ For the definitions of the parameters, see equations (\ref{eq:fermi_dirac}), (\ref{eq:phabs}), (\ref{eq:gauss}), (\ref{eq:spin_phase_model}), and (\ref{eq:gabs}). $^{(b)}$ The equivalent width of the Fe line. $^{(c)}$ 4--78 keV absorbed flux.}
\end{deluxetable*}

%% file: section5.tex
\section{Doubly-phase-resolved analysis}\label{sec:independence}

In Sections \ref{sec:orbital_phase_resolved} and \ref{sec:spin_phase_resolved}, we have investigated the orbital- and spin-phase variability, respectively.
In this section, we examine the orbital- and spin-phase variability simultaneously.
This analysis was carried out to support the validity of the orbital- and spin-phase-resolved analysis by showing that these two variabilities do not affect each other.
We examined the relation between these two ways of variability through light curves and pulse profiles for multiple energy bands.
Since the light curve represents the orbital-phase variability, we resolved it by the spin phase to see both orbital- and spin-phase variability.
Similarly, as the pulse profile represents the spin-phase variability, we resolved it by the orbital phase.

\subsection{Energy- and spin-phase-resolved light curve}\label{sec:fine_light_curve}

We generated light curves resolved by both energy and the spin phase.
By analyzing them, one can examine the dependency of orbital-phase variability on the spin phase.
For the energy, the \nustar\ 3--78 keV band was divided into four energy bands, which are 3--5 keV, 5--10 keV, 10--20 keV, and 20--78 keV.
For the spin phase, we employed the spin-phase intervals A--E defined in Section \ref{sec:spin_ratio_spectrum} and divided the whole spin phase into five intervals.
In order to compare the spin-phase variation at each orbital phase, we define the normalized light curve by
\begin{eqnarray}
L_{\rm norm}(\Phi;\;E_i,\;\phi_j)=\frac{L(\Phi;\;E_i,\;\phi_j)}{\frac{1}{T}\int_{\Phi}L(\Phi;\;E_i,\;\phi_j)d\Phi},
\end{eqnarray}
where $L(\Phi;\;E_i,\;\phi_j)$ denotes the light curve function at the energy band $E_i$ and spin phase $\phi_j$, and $T$ is the net exposure time.
Normalizing each light curve with its average value makes it easy to compare the spin-phase deviation of light curves at each orbital phase.

\begin{figure}[htb!]
\centering
\includegraphics[width=240pt]{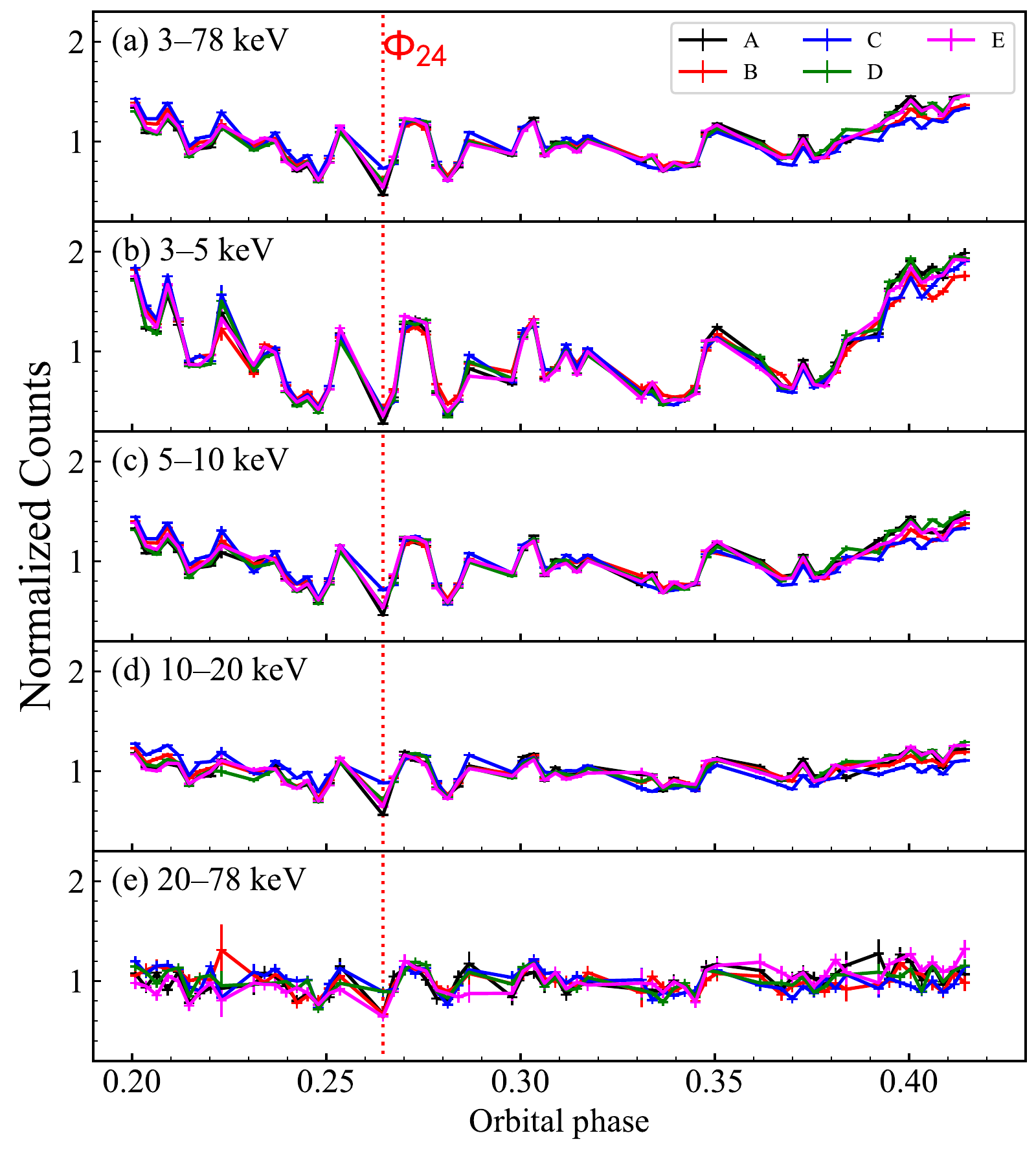}
\caption{Energy- and spin-phase-resolved light curves normalized by their average values (see text). Each panel corresponds to an energy band of (a) 3--78 keV, (b) 3--5 keV, (c) 5--10 keV, (d) 10--20 keV, and (e) 20--78 keV, respectively. Different colors represent different spin phases, A (black), B (red), C (blue), D (green), and E (magenta). See Section \ref{sec:spin_ratio_spectrum} for the definition of spin phase A--E. The orbital phase with most light curve variation ($\Phi_{24}$) is plotted by red dashed line.}
\label{fig:fine_light_curve}
\end{figure}

Figure \ref{fig:fine_light_curve} shows the normalized light curves for multiple energy bands and spin phases.
The difference among spin-phase-resolved light curves at a specific energy band can be evaluated from it.
For every energy band, the spin-phase deviations of light curves appear small because light curves with different colors follow almost the same trajectories.
One can evaluate spin-phase variation of light curves at a given orbital phase $\Phi_i$ and energy band $E_j$ by

\begin{scriptsize}
\begin{eqnarray}
&\Delta& L_{\rm norm}(\Phi_i;\;E_j)\nonumber\\
&=&\frac{\max_{\phi\in A-E}|L_{\rm norm}(\Phi_i;\;E_j,\;\phi)-L_{\rm norm, ave}(\Phi_i;\;E_j)|}{L_{\rm norm, ave}(\Phi_i;\;E_j)},
\end{eqnarray}
\end{scriptsize}
where
\begin{eqnarray}
L_{\rm norm, ave}(\Phi_i;\;E_j)=\frac{1}{5}\sum_{\phi\in A-E}L_{\rm norm}(\Phi_i;\;E_j,\;\phi).
\end{eqnarray}
The average of $\Delta L_{\rm norm}$ is $6.1\%$, $7.9\%$, $6.2\%$, $7.2\%$, and $10.8\%$ for 3--78 keV, 3--5 keV, 5--10 keV, 10--20 keV, and 20--78 keV, respectively.
Therefore, we can conclude that the light curves of different spin phases track very similar variability with variation rates of $\lesssim10\%$.
The 24th orbital interval ($\Phi_{24}=0.265$, red dashed line in Figure \ref{fig:fine_light_curve}) might be an exceptional orbital phase.
It has the most prominent $\Delta L_{\rm norm}$ in all the energy bands, which means it is an orbital phase with the most spin-phase variation.
At this orbital phase, $\Delta L_{\rm norm}$ is $25\%$, $25\%$, $27\%$, $26\%$, and $36\%$ for 3--78 keV, 3--5 keV, 5--10 keV, 10--20 keV, and 20--78 keV, respectively.
Despite the extremely high variability along the spin phase, the orbital-phase-resolved spectrum of this interval does not show any characteristic features compared to those of other orbital intervals.

\subsection{Energy- and orbital-phase-resolved pulse profile}\label{sec:fine_pulse_profile}

Similar to Section \ref{sec:fine_light_curve}, one can examine the dependency of spin-phase variability on the orbital phase by analyzing orbital-phase-resolved pulse profiles.
We generated pulse profiles that were resolved by energy and orbital phase.
For the energy, we adopted the same four energy bands as those of the light curve analysis, 3--5 keV, 5--10 keV, 10--20 keV, and 20--78 keV.
For the orbital phase, we adopted the same orbital intervals used in Section \ref{sec:orbital_phase_resolved}, defined by $\Phi_i$ ($1\leq i\leq78$).
In order to evaluate the variability of pulse profiles, we employed four barometers that characterize the shape of pulse profiles.
One is the rms pulse fraction ${\rm PF_{rms}}$, defined by equation (\ref{eq:rms_pulse_fraction}).
The other three are the ratio among Fourier coefficients, defined in equation (\ref{eq:fourier_coeffcients_sum}).
They are 1st to 0-th Fourier coefficient ($c_1/c_0$), 2nd to 0-th Fourier coefficient ($c_2/c_0$), and 2nd to 1st Fourier coefficient ($c_2/c_1$).

\begin{figure}[htb!]
\centering
\includegraphics[width=240pt]{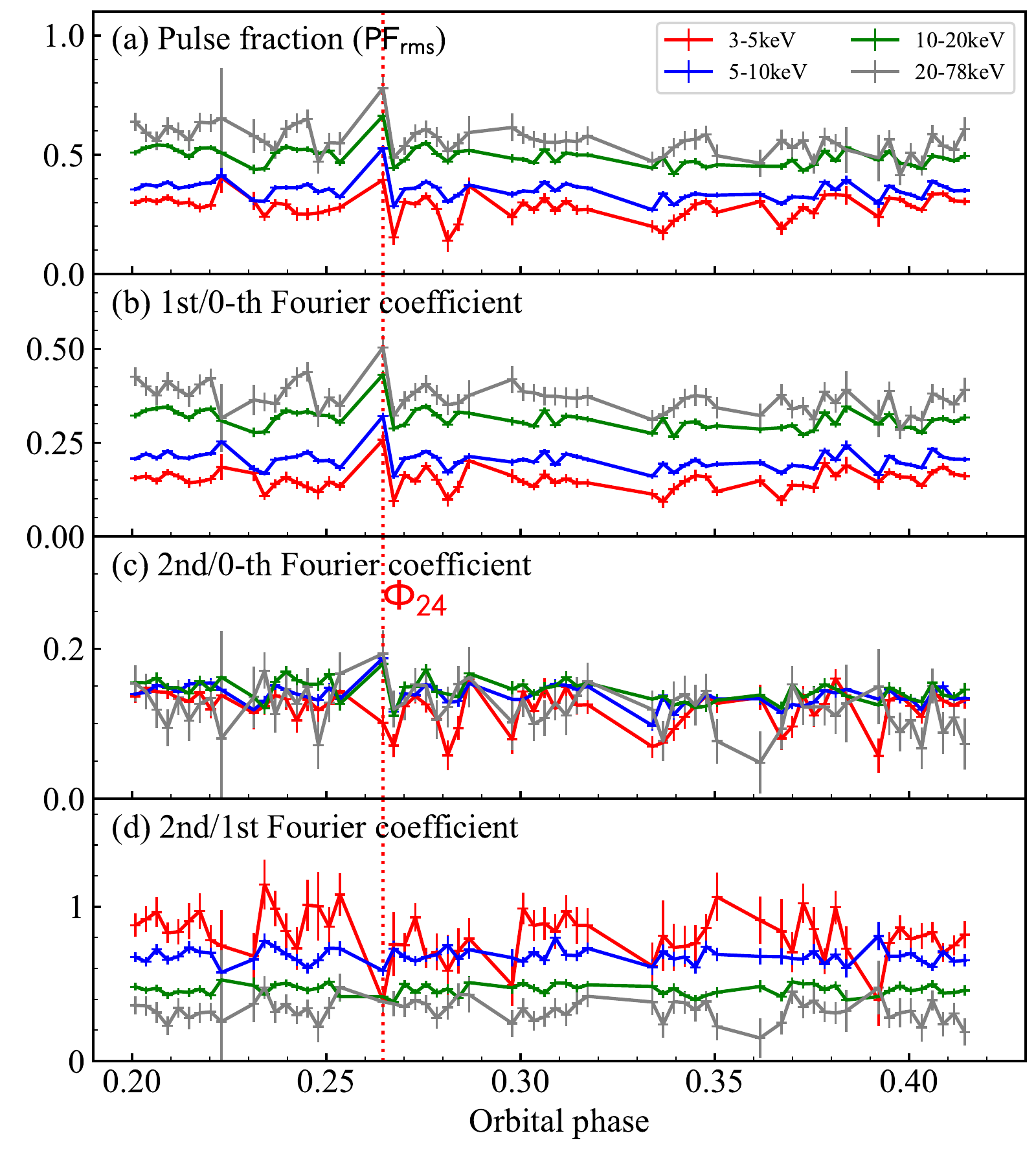}
\caption{Parameters of energy- and orbital-phase-resolved pulse profiles. Each panel denotes (a) Pulse fraction (${\rm PF_{rms}}$), (b) ratio of 1st/0th Fourier coefficient ($c_1/c_0$), (c) 2nd/0th ($c_2/c_0$), and (d) 2nd/1st ($c_2/c_0$), respectively. Different colors represent different energy bands, 3--5 keV (red), 5--10 keV (blue), 10--20 keV (green), and 20--78 keV (gray). The orbital phase with most pulse fraction ($\Phi_{24}$) is plotted by red dashed line.}
\label{fig:fine_pulse_profile}
\end{figure}

Figure \ref{fig:fine_pulse_profile} presents the orbital-phase variability of the four parameters, ${\rm PF_{rms}}$, $c_1/c_0$, $c_2/c_0$, and $c_2/c_1$ for the four energy bands.
Statistical parameters that denote orbital-phase variability, such as the average value (mean), standard deviation ($\sigma$), variation degree ($\sigma$/mean), and reduced chi-square value with respect to the best-fit constant model, are shown in Table \ref{tab:fine_pulse_profile}.
Referring to the variation degrees and reduced chi-square values, there were significant variations in ${\rm PF_{rms}}$, $c_1/c_0$, $c_2/c_0$, while we did not detect significant variations in $c_2/c_1$.
The variation degrees of the 1st and 2nd Fourier components were comparable at all the energy bands.
The variation degree of the ${\rm PF_{rms}}$ above $5\;{\rm keV}$ was roughly $\sim10\%$, which is consistent with the average light curve variability of $5$--$10\%$ along the spin phase derived in Section \ref{sec:fine_light_curve}.
For 3--5 keV, the variation degrees were $\sim20\%$, which is significantly larger than the other bands.
From these results, we conclude that the pulse profiles are highly stable along the orbital phase with variation degrees of $10\%$ for $>5\;{\rm keV}$ and $\sim20\%$ for 3--5 keV.
$\Phi_{24}$ (red dashed line in Figure \ref{fig:fine_pulse_profile}) might be an exception because it shows extremely high pulse fractions compared to other orbital intervals, which is consistent with the spin-phase-resolved light curve analysis (Section \ref{sec:fine_light_curve}).

\begin{deluxetable*}{cccccc}
\tablecaption{Statistical parameters of energy and orbital-phase-resolved pulse profiles.\label{tab:fine_pulse_profile}}
\tablewidth{0pt}
\tablehead{
\colhead{} & \colhead{band} & \colhead{mean} & \colhead{$\sigma^{a}$} & \colhead{$\sigma$/mean} & \colhead{$\chi_\nu^2$ $^{b}$}
}
\startdata
\multirow{4}{*}{${\rm PF_{rms}}$} & 3--5 keV & $0.281\pm0.003$ & $0.050\pm0.002$ & $0.179\pm0.008$ & 3.66\\
& 5--10 keV & $0.351\pm0.001$ & $0.038\pm0.0009$ & $0.108\pm0.003$ & 10.85\\
& 10--20 keV & $0.492\pm0.002$ & $0.039\pm0.003$ & $0.079\pm0.005$ & 8.89\\
& 20--78 keV & $0.559\pm0.007$ & $0.059\pm0.030$ & $0.106\pm0.06$ & 1.72\\
\hline
\multirow{4}{*}{$c_1/c_0$}& 3--5 keV & $0.150\pm0.002$ & $0.028\pm0.0004$ & $0.186\pm0.003$ & 3.46\\
& 5--10 keV & $0.204\pm0.001$ & $0.025\pm0.0002$ & $0.122\pm0.001$ & 9.63\\
& 10--20 keV & $0.313\pm0.001$ & $0.026\pm0.0007$ & $0.083\pm0.002$ & 8.27\\
& 20--78 keV & $0.369\pm0.004$ & $0.037\pm0.006$ & $0.102\pm0.02$ & 1.77\\
\hline
\multirow{4}{*}{$c_2/c_0$}& 3--5 keV & $0.121\pm0.002$ & $0.024\pm0.0003$ & $0.197\pm0.004$ & 2.60\\
& 5--10 keV & $0.138\pm0.001$ & $0.014\pm0.0002$ & $0.098\pm0.001$ & 3.61\\
& 10--20 keV & $0.145\pm0.001$ & $0.014\pm0.0003$ & $0.098\pm0.002$ & 2.81\\
& 20--78 keV & $0.122\pm0.004$ & $0.028\pm0.001$ & $0.233\pm0.01$ & 0.98\\
\hline
\multirow{4}{*}{$c_2/c_1$}& 3--5 keV & $0.819\pm0.02$ & $0.149\pm0.2$ & $0.18\pm0.2$ & 1.59\\
& 5--10 keV & $0.681\pm0.006$ & $0.048\pm0.04$ & $0.071\pm0.06$ & 1.25\\
& 10--20 keV & $0.463\pm0.004$ & $0.034\pm0.01$ & $0.073\pm0.03$ & 1.27\\
& 20--78 keV & $0.331\pm0.01$ & $0.071\pm0.03$ & $0.215\pm0.09$ & 0.72\\
\enddata
\tablecomments{$^{(a)}$ The standard deviation of the parameter. $^{(b)}$ Reduced chi-square when fitted by a constant model.}
\end{deluxetable*}

%% file: section6.tex
\section{Discussion}\label{sec:discussion}

\subsection{Inhomogeneous stellar wind}\label{sec:inhomogeneous_stellar_wind}

In Section \ref{sec:orbital_phase_resolved}, we revealed that a large part of orbital-phase spectral variability was caused by different absorption degrees of stellar wind, especially in low energy bands.
The inhomogeneity of the stellar wind is often interpreted as a result of a clumpy stellar wind.
As seen in Figure \ref{fig:orbital_spectral_parameters}, there are multiple peaks in the variation of $\nh$.
The typical interval of each peak is approximately $\Delta\Phi=0.01$, which corresponds to a timescale of $\sim2\;{\rm ks}$.
As the orbital velocity of the neutron star is $4.4\times10^{7}\;{\rm cm/s}$, the typical size of clumps is
\begin{eqnarray}
R_{\rm c}\sim9\times10^{10}\;{\rm cm}.
\end{eqnarray}
The maximum value of $\nh$ is $\sim3\times10^{23}\;{\rm cm^{-2}}$ as can been seen in Figure \ref{fig:flux_to_nh}.
Therefore, the maximum number density of stellar wind clumps can be estimated to be
\begin{eqnarray}
n_{\rm c}\sim3\times10^{12}\;{\rm cm^{-3}}.
\end{eqnarray}

The typical size and number density of the clumps are reasonable compared to the estimation for Vela~X-1, which is a wind-fed X-ray pulsar with a similar optical companion to Cen~X-3.
\cite{Odaka2013} has estimated the maximum size of the clumps as $R_{\rm c}\sim2\times10^{11}\;{\rm cm}$, which is comparable to Cen~X-3.
They also calculated the typical number density of the clumps as $n_{\rm c}\sim7\times10^{11}\;{\rm cm^{-3}}$.
While $R_{\rm c}$ is comparable for the two sources, $n_{\rm c}$ of Cen~X-3 is larger than that of Vela~X-1 by a factor of 5.
Such a difference can be naturally generated by the interaction between the moving neutron star and stellar wind
\citep[e.g.,][]{Friend1982, Blondin1991}.


\subsection{Stability of accretion stream}\label{sec:stability_of_accretion_stream}

In Section \ref{sec:orbital_phase_resolved}, we performed orbital-phase-resolved spectroscopy and concluded that the flux variability caused by stellar wind absorption is comparable to that caused by intrinsic factors.
The intrinsic factors can be attributed to the variability of the accretion stream.
The 4--78 keV unabsorbed flux variability is only $\sim10\%$, and the accretion stream is rather stable.
Moreover, the spin-phase variability pattern also shows certain stability, as presented in Section \ref{sec:independence}.
The light curves with different spin phases show variations within only $\sim10\%$ (Section \ref{sec:fine_light_curve}), and pulse profiles are stable along the orbital phase within $\sim20\%$ variations at all the energy bands (Section \ref{sec:fine_pulse_profile}).

The possibility of stable accretion flow in Cen~X-3 was also suggested by several previous studies.
A long-term observation via the light curve obtained by the all-sky monitor onboard \ginga\ revealed that there is no correlation between X-ray intensity and the spin-up rate of the spin period.
It suggests that the variation of the light curve does not represent variation in mass accretion rate \citep{Tsunemi1996}.
A long-term light curve observation with the all-sky monitor onboard \rxte\ also supported the possibility. According to the 5--12 keV light curve observed for ten years, there appears to be a ceiling in the maximum luminosity that may correspond to the Eddington luminosity \citep{Paul2005, Raichur2008}.
Moreover, \cite{Naik2011}, which analyzed observation data of Suzaku that cover a full orbit of Cen~X-3, pointed out that the flux decreases (dips) emerging at out-of-eclipse phases are caused by the obscuration of the X-ray source by clumps of dense matter along the line of sight.
All of these results are strong evidence for a stable accretion stream.

Some previous observations suggest inevitable variations in the accretion rate.
According to an \xmm\ observation on the orbital phase of $\Phi=0.36$--$0.80$, the count ratio between 3--10 keV and 0.2--3 keV does not show any significant variance, which suggests that the flux variability originated from the change of mass accretion rate rather than absorption \citep{Sanjurjo2021}.
Although we found an apparent variability in the hardness ratio with the NuSTAR data (Figure \ref{fig:lc_and_hr}b), the invariant hardness ratio in the \xmm\ data may be due to the difference in the time scale of orbital intervals.
The duration of each dip in the \xmm\ data are $<100\;{\rm s}$, a much shorter time scale than the dips observed in the NuSTAR data ($\sim2\;{\rm ks}$).
The variability of the accretion rate may be induced with such a short time scale.
Indeed, we derived the intrinsic flux variability as $\sim10\%$, and other parameters than $\nh$ do not vary significantly along the orbital phase (see \ref{sec:orbital_spectral_parameters}).
This feature is very similar to the \xmm\ observation, where the spectra of the low-flux phase (dip) and high-flux phase (out-of-dip) showed no difference in parameters except for the flux variation of $\sim10\%$ \citep[see Table 5 of][]{Sanjurjo2021}.


\subsection{Spin-phase variability of spectral components}\label{sec:discussion2}
In Section \ref{sec:spin_spectral_parameters}, we performed spin-phase-resolved spectroscopy and demonstrated that each resolved spectrum could be reproduced by a Fermi-Dirac cut-off power law accompanied by an Fe emission line and CRSF.
Here we discuss the spin-phase variability of each spectral component.

\subsubsection{Continuum}\label{sec:discussion_continuum}

The continuum spectrum reproduced by the Fermi-Dirac cut-off power law is characterized by the photon index $\Gamma$, the cut-off energy $E_{\rm c}$, and the folding energy $E_{\rm f}$.
Among the three parameters, $E_{\rm c}$ and $E_{\rm f}$ show no significant variation along the spin phase, as shown in Table \ref{tab:spin_phase_fitting_parameters}.
$\Gamma$, on the other hand, is significantly varied and negatively correlated to the absorbed flux (see Figure \ref{fig:spin_phase_parameter_correlation}a).
The harder photon index in higher-flux phases can be interpreted as a consequence of different optical depths of Comptonization.

To quantify the Comptonization effects, we performed a simplified fitting for 7.5--15 keV spectra of the spin phase A--E.
The energy band is selected because we need to check the pure contribution of Comptonization and exclude the effects of the Fe line, CRSF, and the cut-off.
The employed model is {\tt compTT} in {\tt XSPEC}, which is an analytic model describing the Comptonization of soft photons in a hot plasma, developed by \cite{Titarchuk1994}.
The input soft photon temperature was fixed to a sufficiently small value of $0.1\;{\rm keV}$, and the plasma temperature was also fixed to $4.77\;{\rm keV}$.
The plasma temperature is suggested for the accretion column of Cen~X-3 by \cite{Thalhammer2021}, and very close to $E_{\rm f}$ values in Table \ref{tab:spin_phase_fitting_parameters}.
Note that the folding energy is related to the electron temperature in the case of unsaturated Comptonization \citep[e.g.,][]{Rybicki1979}.
The only free parameters are the optical depth of Comptonization $\tau$ and the normalization factor.
The fitting results yielded $\tau=7.7$, $9.2$, $9.4$, $12.4$, and $8.2$ for spin-phase intervals A, B, C, D, and E, respectively.
The optical depth is strongly correlated with the photon index, with a harder photon index corresponding to a larger optical depth.
The difference between the maximum and minimum value of optical depth is a factor of $\sim1.6$.

\subsubsection{Fe emission line}

As presented in Figure \ref{fig:spin_phase_parameter_correlation}(c), the spin-phase-resolved spectroscopy has revealed that the equivalent width of the Fe emission line is negatively correlated to the flux.
This is natural for X-ray pulsars because a part of Fe lines should be consequences of X-ray illumination of stellar wind, originating from non-pulsed regions apart from the neutron star.
The orbital-phase variations of Fe emission lines examined by previous researches support this assumption, as the equivalent width of Fe emission lines significantly increases in dip and eclipse phases \citep{Ebisawa1996, Naik2011}.
\red{The decreased pulse fraction in 6--7 keV as shown in Figure \ref{fig:pulse_profile_analysis}(a) also reflects the lower spin-phase variation of the Fe line.
Such energy dependence of pulse fraction was also seen in a transient X-ray pulsar GRO~J1744-28, where the pulse fraction of 6--7 keV is lower than nearby bands \cite{Nishiuchi1999}.}

In Figure \ref{fig:spin_phase_parameter_correlation}(b), we also see a negative correlation between the intensity of the Fe line and the flux.
The spin-phase variability of the Fe line intensity strongly suggests that there is also line emission from the proximity of the neutron star, like the accretion disk or accretion flow.
However, the negative correlation between the line intensity and the continuum flux is not quite simple to interpret.
Similar trends were reported by spin-phase-resolved spectroscopy on Ginga data \citep{Day1993} and \bepposax\ data \citep{Burderi2000}.
A possible explanation for the negative correlation would be the time lag of the arrival photons or the geometrical effect such as obscuration by the accretion disk or stream, but there is still no convincing observational evidence.
This is partly because of large uncertainties of the Fe line measurement since it is sensitive to the modeling of the continuum and could easily couple with $\nh$.
It is also necessary to distinguish the contributions from neutral and highly ionized Fe emission lines.
A precise measurement should be done by a detector with a good time and energy resolution, such as XRISM \citep{Tashiro2018}.


\subsubsection{Cyclotron resonance scattering feature}

As presented in Figure \ref{fig:spin_phase_parameter_correlation}(d), the central energy of the CRSF ($E_{\rm cyc}$) shows a significant variation along the spin phase and a positive correlation with the flux.
$E_{\rm cyc}=28.7\;{\rm keV}$ at the pulse maximum is a consistent value with those detected by other satellites \citep[][also see Table 4 of \citealp{Tomar2021}]{Nagase1992, Santangelo1998, Burderi2000, Suchy2008}.
The minimum $E_{\rm cyc}$ of $26.0\;{\rm keV}$ and maximum $E_{\rm cyc}$ of $28.7\;{\rm keV}$ correspond to magnetic fields of $B=2.2\times10^{12}\;{\rm G}$ and $2.5\times10^{12}\;{\rm G}$, respectively.
It means that the average magnetic field that photons experience is different depending on the emission direction by $\sim10\%$.
This result is qualitatively consistent with the spin-phase-resolved spectroscopy on \bepposax\ data \citep{Burderi2000}.
Their analysis yielded the highest central energy of CRSF at the pulse peak and lowest in the pulse minimum but pointed out that $E_{\rm cyc}$ ranges from $28$ to $36\;{\rm keV}$, which is a much larger variation than our results.
The disagreement is presumably due to the different modeling of the continuum as the CRSF is quite sensitive to how we evaluate the continuum spectrum.
The energy-resolved pulse fraction is another evidence that the CRSF center is in the range of 25--30 keV since it has a dented feature at this energy band and a much larger pulse fraction at 30--35 keV (see Figure \ref{fig:pulse_profile_analysis}a).
\red{The positive correlation of the CRSF energy and spin-phase flux was also reported for the X-ray pulsar V~0332+53 \citep{Lutovinov2015}.
It displays a similar CRSF energy shift to Cen~X-3 ranging from 23 to 27 keV, which can be explained by the reflection from the neutron star surface \citep{Poutanen2013}.
A similar interpretation can be applied to Cen~X-3: the reflected photons by the proximity of the magnetic pole result in the large $E_{\rm cyc}$ in the pulse maximum, while those reflected by distant region from the pole result in the small $E_{\rm cyc}$ in the pulse minimum.}

Figure \ref{fig:spin_phase_parameter_correlation}(e) shows a significant positive correlation between the standard deviation of CRSF ($\sigma_{\rm cyc}$) with the flux.
The optical depth at the central energy of CRSF ($\tau_{\rm cyc}$) also seems to show a slight variation along the spin phase (Figure \ref{fig:spin_phase_parameter_correlation}f), although it is not significant enough.
These two parameters represent the strength of CRSF, which is a barometer to measure how long the photons traveled in the dense plasma under strong magnetic fields.
The strength of CRSF differs by a factor of $1.9\pm0.5$, which is calculated from $\sigma_{\rm cyc}$ and $\tau_{\rm cyc}$ at the pulse maximum and minimum.
This value is roughly consistent with the difference in optical depth of Comptonization, which is $\sim1.6$ as calculated in Section \ref{sec:discussion_continuum}.
This is reasonable because the location where the CRSF is formed is the same as that of Comptonization, both of which take place in the optically thick plasma with strong magnetic fields.

\subsection{Multiple components of pulsed emission}\label{sec:multiple_components_of_emission}

\begin{figure*}[htb!]
\centering
\includegraphics[width=\linewidth]{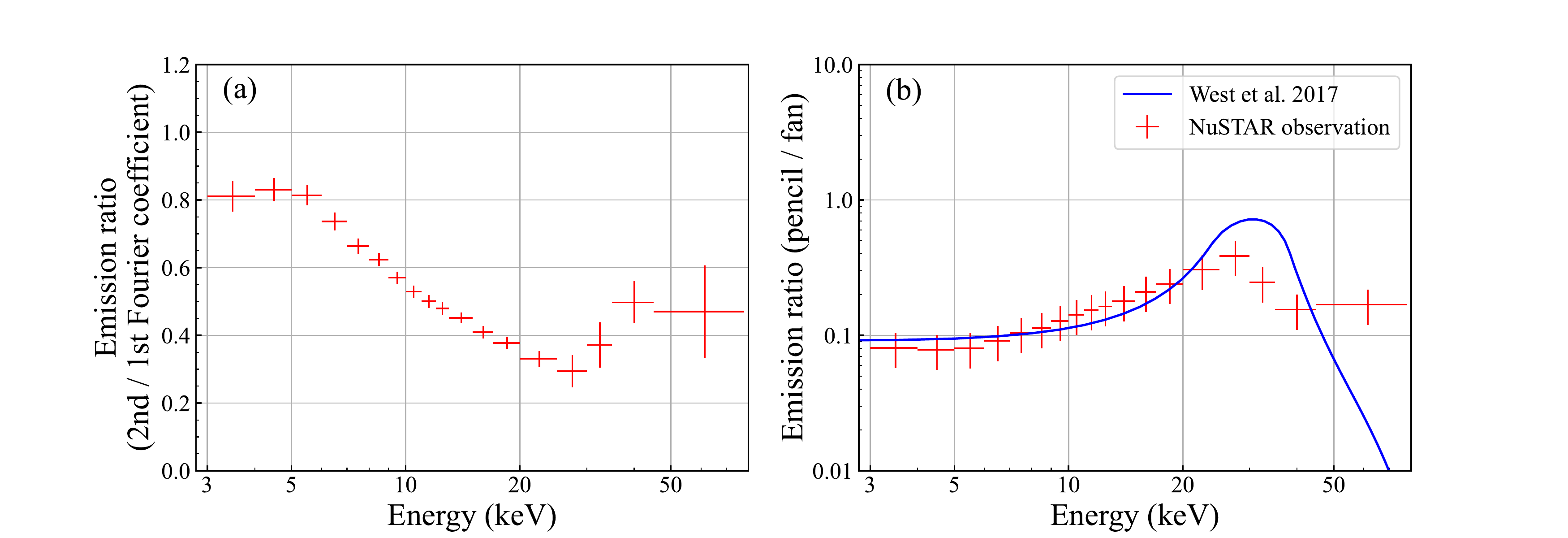}
\caption{(a) Emission ratio of 2nd to 1st Fourier coefficient plotted as a function of energy, derived from the \nustar\ observation. (b) Emission ratio of the pencil beam to the fan beam as a function of energy. The red plots denote ratio calculated from the observation data with the best-fit parameter set in equation (\ref{eq:pencil_fan_best_parameters}). The blue line represents the simulation-based calculation by \cite{West2017}.}
\label{fig:pencil_fan}
\end{figure*}

In Section \ref{sec:pulse_profile}, we investigated energy-resolved pulse profiles and derived the contributions of the 1st and 2nd Fourier components at each energy band.
The outcome gives us crucial hints about emission geometries around the neutron star.
In Figure \ref{fig:pencil_fan}(a), we presented the energy dependence of the ratio of the 2nd Fourier coefficient to the 1st Fourier coefficient ($c_2/c_1$).
The ratio is lowest at 25--30 keV, which means the 1st Fourier component is most dominant at this energy band.
This property is similar to a simulation-based calculation by \cite{West2017} applied to \rxte\ observation data.
In the calculation, they assumed the accretion column model \citep{Becker2007} and derived the X-ray emission spectrum from the top of the column (pencil beam) as well as that from the wall of the column (fan beam) for several X-ray pulsars.
According to the calculation, the pencil-to-fan ratio peaks at $\sim 30\;{\rm keV}$ in the case of Cen~X-3 \citep[see Figure 11 of][]{West2017}.

We performed a simple calculation that links the Fourier coefficients to the pencil and fan beam flux.
We assumed that the emission from the neutron star is composed of two components, pencil and fan beam, and certain parts of them are observed as the 1st and 2nd Fourier components.
Specifically, we assume
\begin{eqnarray}
a_1F_{\rm pen}+b_1F_{\rm fan}&=&F_{\rm 1st}\\
a_2F_{\rm pen}+b_2F_{\rm fan}&=&F_{\rm 2nd},
\label{eq:pencil_fan_best_parameters}
\end{eqnarray}
where $F_{\rm pen}$, $F_{\rm fan}$ are the fluxes of pencil and fan beam, and $F_{\rm 1st}$ and $F_{\rm 2nd}$ are the fluxes of 1st and 2nd Fourier component, respectively.
$a_1$, $a_2$, $b_1$, $b_2$ are non-negative coefficients that satisfy $0\leq a_1+a_2\leq1$ and $0\leq b_1+b_2\leq1$.

We performed a chi-squared fitting by setting the four coefficients as varying parameters, and searched for the parameter set where the observation data agreed with the simulation-based calculation best.
For $F_{\rm 1st}$ and $F_{\rm 2nd}$, we adopted $c_1$ and $c_2$ as normalized values, which are the 1st and 2nd Fourier coefficients derived from the pulse profile analysis (Section \ref{sec:pulse_profile}).
The best-fit parameters were
\begin{eqnarray}
a_1&=&0.86,\;a_2=0.14,\\
b_1&=&0.00,\;b_2=0.05.
\end{eqnarray}
The fitting result is presented in Figure \ref{fig:pencil_fan}(b).
Since we only compare the ratio between the emission components, multiplying a common constant to the four parameters does not affect the fitting result.
Therefore, the we adopted $a_1+a_2=1$ to set the normalization of the parameters for simplicity.
The yielded best-fit parameters suggest that the 1st Fourier component is more linked to the pencil beam while the 2nd is to the fan beam.
This is natural when we assume two antipodal emission regions because the pencil beam can be interpreted as single-peaked emission while the fan beam as double-peaked emission.
The fitting result also suggests that a large part of non-pulsed component is generated from the fan beam.
The remained small discrepancy between the observation and the theoretical calculation may be due to the simplified linear transformation, but the precise estimation of the emission components is beyond the scope of this paper.
\red{A possible effect that should be considered is the displacement from the dipole geometry by $\sim10^{\circ}$, which is suggested by the pulse profile decomposition \citep{Kraus1996} and the spin-phase variability of polarization observed by IXPE \citep{Tsygankov2022}.}

%% file: section7.tex
\section{Conclusions}\label{sec:conclusions}

We studied the orbital- and spin-phase variability of the X-ray pulsar Cen~X-3 by analyzing $39\;{\rm ks}$ \nustar\ observation data covering an orbital-phase interval of $\Phi=0.199$--$0.414$.
The orbital-phase variability was a consequence of the mixture of two comparable effects, intrinsic flux variability of $\sim10\%$ and the obscuration by the inhomogeneous stellar wind with a maximum column density of $\nh\sim3\times10^{23}\;{\rm cm^{-2}}$.
The measured column density and time scale of dips indicate that the typical size and density of clumps are $R_{\rm c}\sim9\times10^{10}\;{\rm cm}$ and $n_{\rm c}\sim3\times10^{12}\;{\rm cm^{-3}}$, respectively.
The spin-phase variability was caused by the difference in the optical depth of Comptonization, estimated from the significant hardening of the photon index.
The photon index variation of $0.72$--$1.06$ corresponds to the optical depths of $7.7$--$12.4$, a difference by a factor of $\sim1.6$.
The spectral features of the CRSF and the Fe line also showed variability along the spin phase.
The CRSF variability was characterized by the enhancement in the central energy and the optical depth at the pulse maximum, where the central energy ranged from 26.0 keV to 28.7 keV and the optical depth varied by a factor of $\sim 1.9$.
The variation degrees of the optical depth of the Comptonization and CRSF were comparable, implying these two effects originate from the same region.
The Fe line variability was represented by the decrease in the Fe line flux at the pulse maximum, which possibly originates from the time lag of photon travel time or geometrical effects of the accretion disk and stream.
The pulse profile was double-peaked in the low-energy band and gradually shifted to single-peaked with energy.
This supports the existence of two different emission patterns corresponding to the pencil and fan beams.
Finally, we found that the pulse profiles were stable along the orbital phase within a variation degree of $\sim20\%$, which gives evidence of a highly stable accretion stream in the binary system.